\documentclass{emulateapj}

\def\lesssim{\mathrel{\hbox{\rlap{\hbox{\lower4pt\hbox{$\sim$}}}\hbox{$<$}}}}
\def\gtrsim{\mathrel{\hbox{\rlap{\hbox{\lower4pt\hbox{$\sim$}}}\hbox{$>$}}}}

\slugcomment{}
\shorttitle{Dissipation and Vertical Energy Transport}
\shortauthors{Blaes et al.}

\begin{document}

\title{Dissipation and Vertical Energy Transport in Radiation-Dominated
Accretion Disks}

\author{Omer Blaes}
\affil{Department of Physics, University of California, Santa Barbara,
   Santa Barbara CA 93106}

\author{Julian H. Krolik}
\affil{Department of Physics and Astronomy, Johns Hopkins University,
    Baltimore, MD 21218}

\author{Shigenobu Hirose}
\affil{Institute for Research on Earth Evolution, JAMSTEC, Yokohama,
Kanagawa 236-0001, Japan}

\and

\author{Natalia Shabaltas\footnote{now at Department of Physics, Cornell
   University, 109 Clark Hall, Ithaca, NY 14853}}
\affil{Department of Physics, University of California, Santa Barbara,
   Santa Barbara CA 93106}

\begin{abstract}
Standard models of radiation supported accretion disks generally assume
that diffusive radiation flux is solely responsible for vertical heat transport.
This requires that heat {\it must} be generated at a critical rate
per unit volume if the disk is to be in hydrostatic and thermal equilibrium.
This raises the question of how heat is generated
and how energy is transported in MHD turbulence.  By analysis of a
number of radiation/MHD stratified shearing-box simulations,
we show that the divergence of the diffusive radiation flux is indeed
capped at the critical rate, but deep inside the disk, substantial
vertical energy flux is also carried by advection of radiation.
Work done by radiation pressure is a significant part of the energy
budget, and much of this work is dissipated later through damping by radiative
diffusion.
We show how this damping can be measured in the simulations, and identify
its physical origins.  Radiative damping accounts for as much as tens of percent
of the total dissipation,  and is the only realistic physical mechanism for
dissipation of turbulence that can actually be resolved in numerical
simulations of accretion disks.  Buoyancy associated
with dynamo-driven, highly magnetized, nearly-isobaric nonlinear slow
magnetosonic fluctuations is responsible for the radiation advection flux,
and also explains the persistent periodic magnetic upwelling seen at all
values of the radiation to gas pressure ratio.
The intimate connection between radiation advection and magnetic buoyancy
is the first example we know of in astrophysics in which a dynamo has
direct impact on the global energetics of a system.
\end{abstract}


\section{Introduction}
\label{sec:introduction}

Much of the power radiated by the accretion flow in luminous states of X-ray
binaries and active galactic nuclei necessarily originates from the vicinity
of the central compact object where the gravity well is deepest.  If the
bolometric luminosity is anywhere close to Eddington, radiation pressure
must dominate the thermal pressure of the accreting plasma in these regions,
and understanding the physics of radiation dominated accretion is therefore
central to any explanation of how these sources work in their brightest
states.  Classical accretion disk theory \citep{sha73,nov73} models the flow
as geometrically thin and optically thick, and assumes that angular
momentum is transported within the flow by internal stresses due to turbulence.
The average internal stress is postulated to scale with thermal,
and therefore mostly radiation, pressure.  Accretion power is assumed to be
locally dissipated as heat, which is then transported vertically outward
by photon diffusion.

This model has been questioned over the years on a number of grounds.  First
and foremost, if one takes its assumptions literally, then the resulting
equilibrium structure is unstable to both thermal and inflow (``viscous'')
instabilities \citep{le74,shi75,sun75,sha76}.  However, these assumptions
may not be valid.  \citet{bis77} point out that if the turbulent dissipation
rate per unit mass is constant with height in the radiation pressure dominated
regime, then the mass density would be independent of height inside the
photospheres, a situation that would clearly be convectively
unstable.\footnote{This assumption of constant dissipation per unit mass is
implicit in equation (2.22) of \citet{sha73}, where they go beyond one-zone
modeling and attempt a detailed treatment of the disk vertical structure.  As
a result, they conclude after their equation (2.26) that the density is
independent of height.  It has also been adopted by other authors even in
recent years; e.g. appendix B of \citet{beg06} makes the same assumption in
deriving a simple model of convective transport in the presence of photon
bubbles.}  The argument is simple and worth repeating.  Hydrostatic
equilibrium in a radiation dominated disk implies that the vertial profile of
radiation flux $F(z)$ with height $z$ obeys $\kappa F/c=\Omega^2 z$, where
$\kappa$ is the
opacity (which is very nearly constant as Thomson scattering dominates the
flux mean), $c$ is the speed of light, $\Omega$ is the local angular velocity,
and we have assumed Newtonian gravity for simplicity.  If radiative diffusion
dominates the heat transport, then equilibrium requires that the flux
divergence $dF/dz$ be equal to the dissipation rate per unit volume $Q$,
giving $Q=c\Omega^2/\kappa\equiv Q^\star$, which is constant.\footnote{This
argument also led \cite{sha76} to observe that $2c\Omega/(3\kappa)$
is a characteristic value for the stress in radiation dominated disks.}
If the dissipation rate per unit {\it mass} is also assumed to be constant,
then the density must be constant, implying convective instability.  It should
be noted, however, that vertical energy transport by convection under the
conditions of constant dissipation per unit mass need not alter the thermal
instability of the disk, as diffusive transport
of energy could still be dominant \citep{sha78}.

Energy may also be transported vertically in other ways.  Rather than
completely dissipating locally, magnetic energy in the turbulence
could be transported outward by buoyancy in the form of Poynting
flux, perhaps to be dissipated outside the disk photosphere in a corona
\citep{gal79}.  It has even been argued that magnetic buoyancy would limit the
magnetic energy density in a radiation dominated plasma to be at most
the gas pressure, thereby producing an accretion stress that scales with
the gas pressure alone \citep{sak81,ste84,sak89}.  Such a stress would
eliminate the thermal and inflow instabilities that plague the standard
model.

Unfortunately, this work was built on an incomplete foundation because
it lacked an understanding of the physical nature of angular momentum
transport in accretion disks.  One could therefore only guess the vertical
profile of
dissipation per unit mass, and pretend that the disk can be adequately
modelled by a time-averaged, steady-state vertical structure.  Moreover,
arguments concerning the efficacy of magnetic buoyancy had to make assumptions
about the rate of magnetic field generation, as well as the geometry of
magnetic field lines.

Since the discovery of the relevance of magnetorotational (MRI) turbulence
to accretion disks \citep{bal91,haw91,bal98}, it has become possible to explore
these ideas in detail with numerical simulations.  The most relevant
simulations thus far have used a shearing box that incorporates the
vertical tidal gravitational field of the central
object \citep{bra95,sto96}.  The accretion stress in such simulations
arises self-consistently from correlated magnetic and velocity fluctuations
within the turbulence itself.  Using such stratified shearing boxes
with an isothermal equation of state, \citet{mil00} found that the majority
of the magnetic energy generated in MRI turbulence was (numerically)
dissipated locally.  Nevertheless, a quarter of the magnetic energy generated
was vertically transported outward by buoyancy, forming a strongly magnetized
corona outside a weakly magnetized structure near the disk midplane.
\citet{mil00} used an isothermal equation of state, however, and did not
include the possibility of diffusive radiation transport outward along the
vertical temperature gradients.

Inclusion of such transport would also introduce new dissipation physics,
as compressible waves should be damped by radiative diffusion \citep{ago98}.
This process is entirely analogous to the Silk
damping of acoustic perturbations in the early universe \citep{sil67,sil68},
and to the radiative damping of stellar pulsation modes \citep{cox80}.
It is particularly interesting because such dissipation can easily
be resolved numerically in radiation MHD simulations.
This contrasts sharply with our complete inability to
resolve the microscopic scales on which viscous and resistive
dissipation damps fluid and magnetic fluctuations.

Shearing box simulations of MRI turbulence incorporating
thermodynamics and radiation transport have been possible for some time now.
The first
such simulations neglected vertical gravity and explored the properties of
the turbulence in the presence of radiation transport, treated numerically
using flux-limited diffusion \citep{tur02,tur03}.
If photon diffusion is rapid enough, and the
magnetic pressure exceeds the pressure in the gas alone, then MRI turbulence
can become extremely compressible with strong density fluctuations.
These fluctuations are highly dissipative:
net $PdV$ work is done on the plasma over time, indicating an irreversible
conversion of mechanical energy into internal energy \citep{tur02,tur03}.
This result confirmed the suggestion of \citet{ago98}, although
the specific character of the fluctuations being damped was not entirely
clear.

The first radiation MHD shearing box simulation of MRI turbulence with
vertical gravity was published by \citet{tur04}.  The simulation was
radiation pressure dominated, and exhibited no evidence of thermal instability.
Moreover, the time-averaged vertical entropy profile was stable to hydrodynamic
convection.  On the other hand, the simulation was not fully energy-conserving,
nor was it able to incorporate the photospheres within the simulation domain.
Substantial mass loss occurred during the simulation and this was suggested as
a possible cause of the absence of any exponential thermal runaway.  An
attempt was also made to measure the radiative damping using the time-averaged
$PdV$ work, but the result was necessarily uncertain because this work can
also be used to excite vertical mechanical motions when vertical gravity is
present.

The technical issues with the \citet{tur04} simulation were solved by
\citet{hir06}.  Radiation transfer was still treated by flux-limited diffusion,
but a new diffusion solver permitted several improvements: The quasi-periodic
radial boundary conditions appropriate to shearing boxes could be imposed
properly.  Low densities could now be handled, allowing the photospheres to
be incorporated within the simulation domain.  Most importantly, a total
energy scheme was implemented so that grid scale
losses of magnetic and kinetic energy were fully captured as heat in the
gas, and the code accurately conserves energy.  While dissipation is therefore
still numerical, it may nonetheless mimic local energy flow to microscopic
dissipation scales at high wavenumbers in the turbulent cascade.  Radiation
and gas exchange momentum through Thomson scattering and free-free opacity,
and exchange energy through free-free absorption and emission.  Energy
exchange by Compton scattering was later incorporated into the code in
\citet{hir09}.

Using this improved code, we have succeeded in establishing long-lived
thermal equilibria in which gas pressure dominates \citep{hir06},
gas and radiation pressure are comparable \citep{kro07}, and radiation
pressure dominates \citep{hir09}.  The radiation dominated state is
thermally stable, even with no mass loss, and this is
due to the fact that thermal pressure lags turbulent stress on time
scales of order the thermal time \citep{hir09}.  We have also
established that
the stress and total pressure are correlated (see also \citealt{ohs09}), but
only on time scales longer than the thermal time, reviving the question of
inflow instability
\citep{le74,hirbk09}.  Neither the gas density nor the dissipation rate
per unit mass is constant with height in the radiation dominated (or any other)
regime \citep{hir09}, and the time and horizontally-averaged structures
are stable to hydrodynamic convection \citep{kro07}.  However,
the outer layers are generally magnetically supported \citep{hir06},
and exhibit Parker instability dynamics \citep{bla07}.

There remain a number of important questions about the thermodynamics of
radiation dominated accretion disks.  How does one properly calculate the
contribution of radiative damping to the overall dissipation when $PdV$ work can
also be used to excite vertical mechanical motions?  What exactly are the
compressive motions that are being dissipated by radiative diffusion?  What
happens if the local dissipation rate per unit volume exceeds the radiation
pressure dominated hydrostatic value of $Q^\star=c\Omega^2/\kappa$?
What controls the relative shares of radiative diffusion, advection, and
Poynting flux in vertical energy transport, and how does this depend on
radiation to gas pressure ratio and height?  Can there be significant
coherent energy flux in the form of, e.g., vertical acoustic waves, and if
so, how are these waves excited and how much energy do they transport and
dissipate?  Even if the time- and horizontally-averaged vertical structure
is convectively stable, can local and transient buoyancy lead to significant
energy transport?

The goal of this paper is to answer these questions on the basis of
detailed analysis of simulation data.  In section~\ref{sec:simulations},
we provide a brief overview of the radiation dominated simulations we have
analyzed.  In section~\ref{sec:energetics}, we analyze global energetics,
first deriving and discussing the total energy conservation equation
in section~\ref{sec:energy}, and then turning to the first law of thermodynamics
in section~\ref{sec:thermodynamics}.  There we show that work done by pressure
is increasingly important at high radiation to gas pressure ratios, and this
work is associated both with radiative damping and excitation of vertical
mechanical motions.
We identify two important classes of radiation pressure fluctuation in
section~\ref{sec:pradmodes}, and show how radiative damping acts upon them in
section~\ref{sec:silkdamping}.
We then present detailed results on the nature of vertical advective energy
transport in section~\ref{sec:verticaladvection}.  In
section~\ref{sec:discussion}, we discuss how our findings give rise to
a more dynamic view of the thermal physics of a radiation dominated disk.
Finally, we summarize our
conclusions in section~\ref{sec:conclusions}.  We provide some
mathematical background on trapped vertical modes that modulate the mechanical
work and advective energy transport in an appendix.

\section{Simulations}
\label{sec:simulations}

The radiation MHD equations solved in our simulations are discussed by
\citet{hir06} and \citet{hir09}, but we list them here again as they are the
basis for most of the equations we derive elsewhere in the paper.
\begin{equation}
{\partial\rho\over\partial t}+{\nabla}\cdot(\rho{\bf v})=0
\label{eq:cont}
\end{equation}
\begin{eqnarray}
{\partial\over\partial t}(\rho{\bf v})+\nabla\cdot(\rho{\bf v}{\bf v})&=&
-\nabla p-\nabla\cdot{\sf q}+{1\over4\pi}(\nabla\times{\bf B})\times{\bf B}\cr
&&+{\bar{\kappa}^{\rm R}\rho\over c}{\bf F}+{\bf f}_{\rm SB}
\label{eq:gasmom}
\end{eqnarray}
\begin{eqnarray}
{\partial e\over\partial t}+\nabla\cdot(e{\bf v})&=&-p\nabla\cdot{\bf v}
-{\sf q}:\nabla{\bf v}
-(aT^4-E)c\bar{\kappa}^{\rm P}_{\rm ff}\rho\cr
&&-cE\kappa_{\rm es}\rho{4k_{\rm B}
(T-T_{\rm rad})\over m_ec^2}+\tilde{Q}
\label{eq:gasenergy}
\end{eqnarray}
\begin{eqnarray}
{\partial E\over\partial t}+\nabla\cdot(E{\bf v})&=&-{\sf P}:\nabla{\bf v}
+(aT^4-E)c\bar{\kappa}^{\rm P}_{\rm ff}\rho\cr
&&+cE\kappa_{\rm es}\rho{4k_{\rm B}
(T-T_{\rm rad})\over m_ec^2}-\nabla\cdot{\bf F}
\label{eq:radenergy}
\end{eqnarray}
\begin{equation}
{\partial{\bf B}\over\partial t}=\nabla\times({\bf v}\times{\bf B})
\label{eq:induction}
\end{equation}
\begin{equation}
{\bf F}=-{c\lambda\over\bar{\kappa}^{\rm R}\rho}\nabla E
\label{eq:raddiff}
\end{equation}

Here $\rho$ is the density, ${\bf v}$ is the fluid velocity, ${\bf B}$ is the
magnetic field, $p$ is the pressure in the gas, ${\sf q}$ is a diagonal
tensor associated with the artificial bulk viscosity adopted by the code
\citep{sn92}, $e=3p/2$ is the internal energy density in the gas,
$T$ is the gas temperature, $E$ is the radiation energy density, ${\sf P}$ is
the radiation pressure tensor, $T_{\rm rad}\equiv(E/a)^{1/4}$ is the effective
temperature of the radiation, $a$ is the radiation density constant,
${\bf F}$ is the radiation flux, $k_{\rm B}$ is Boltzmann's constant, $m_e$ is
the electron mass, $\kappa_{\rm es}$ is the electron scattering opacity,
$\bar{\kappa}_{\rm ff}^{\rm P}=\bar{\kappa}_{\rm ff}^{\rm P}(\rho,e)$ is
the Planck mean free-free opacity,
$\bar{\kappa}^{\rm R}=\bar{\kappa}^{\rm R}(\rho,e)$ is
the Rosseland mean opacity including electron scattering and free-free
contributions\footnote{Electron scattering dominates the Rosseland mean
opacity in all the simulations considered in this paper, so that
$\bar{\kappa}^{\rm R}=\kappa_{\rm es}$ quite accurately.},
$\lambda$ is a flux limiter equal to 1/3 in the optically
thick limit, and $\tilde{Q}$ is
the dissipation rate per unit volume required to maintain total energy
conservation due to grid-scale losses of magnetic and kinetic energy.
More details on how many of these quantities are actually computed in the
code can be found in \citet{hir06} and \cite{hir09}.

The vector ${\bf f}_{\rm SB}$ represents the gravitational and inertial forces
in the local shearing box frame, which is rotating at fixed angular velocity
$\Omega$ with respect to an inertial reference frame:
\begin{equation}
{\bf f}_{\rm SB}=-2\rho\Omega\hat{\bf z}\times{\bf v}-\rho\nabla\phi,
\label{eq:fsb}
\end{equation}
where $\phi\equiv-3\Omega^2x^2/2+\Omega^2z^2/2$ is the effective gravitational
potential, and $x$, $y$, $z$ are Cartesian spatial coordinates along the radial,
azimuthal, and vertical directions, respectively.

Table \ref{simparams} summarizes most of the numerical parameters of the
six simulations we analyze in this paper, including the surface mass density
$\Sigma$ and scale height $H$ of the initial condition.  Each simulation had
the same angular velocity $\Omega=190$~rad~s$^{-1}$, corresponding to a radius
of $30GM/c^2$ around a 6.62~M$_\odot$ Schwarzschild black hole.  The grids
all had 48 zones in the $x$ (radial) direction, 96 zones in the $y$ (azimuthal)
direction, and 896 zones in the $z$ (vertical) direction.  The simulations
were all initialized with a weak magnetic field in a twisted azimuthal
flux tube geometry in the midplane regions of the domain.  Magnetorotational
turbulence is well-established by ten orbits into the simulation, and we ran
each simulation a further 250 to 600 orbits.  This is much longer than the
thermal time of the simulations, the average of which ranges from
13 orbits for 1112a to 24 orbits for 0519b, as shown in Table~\ref{simprops}.
All the simulations reach approximate thermal equilibria with continued long
time scale fluctuations.  Radiation pressure dominates the thermal pressure
in all these simulations, and the time and box-averaged ratios of radiation
to gas pressure are also listed in Table \ref{simprops}.  More information
about these simulations can be found in \citet{hir09} and \citet{hirbk09}.

\begin{deluxetable*}{ccccc}
\tablewidth{0pt}
\tablecaption{Simulation Parameters \label{simparams}}
\tablehead{
\colhead{Simulation} & \colhead{$\Sigma$} & \colhead{$H$} & \colhead{Duration} &
\colhead{Box Dimensions}\\
 & \colhead{(g cm$^{-2}$)} &  \colhead{(cm)} & \colhead{(orbits)} &
\colhead{($L_x/H\times L_y/H\times L_z/H$)}
}
\startdata
0211b & $5.43\times10^4$ & $5.83\times10^6$ & 264 &
$0.3375\times1.35\times6.3$\\
0519b & $7.48\times10^4$ & $4.37\times10^6$ & 403 &
$0.3375\times1.35\times6.3$\\
1112a & $1.06\times10^5$ & $1.46\times10^6$ & 610 &
$0.45\times1.8\times8.4$\\
1126b & $1.06\times10^5$ & $1.46\times10^6$ & 611 &
$0.45\times1.8\times8.4$\\
0520a & $1.24\times10^5$ & $1.17\times10^6$ & 603 &
$0.54\times2.16\times10.08$\\
0320a & $1.52\times10^5$ & $7.28\times10^5$ & 426 &
$0.6\times2.4\times11.2$\\
\enddata
\end{deluxetable*}

\begin{deluxetable*}{cccccc}
\tablewidth{0pt}
\tablecaption{Summary of Time-Averaged Physical Properties of Simulations
\label{simprops}}
\tablehead{
\colhead{Simulation} &
\colhead{$H_P/H$\tablenotemark{a}} &
\colhead{$t_{\rm thermal}$/orbits\tablenotemark{b}}&
\colhead{$\left<{P_{\rm rad,av}\over P_{\rm gas,av}}\right>$\tablenotemark{c}} &
\colhead{$\left<{\rho_{\rm max}\over\rho_{\rm min}}\right>$\tablenotemark{d}}
}
\startdata
0211b & 0.55 & $16\pm3.2$ & $62\pm16$ & $2.3\pm0.56$ \\
0519b & 0.81 & $24\pm5.2$ & $70\pm21$ & $2.6\pm1.3$ \\
1112a & 0.71 & $13\pm2.6$ & $6.6\pm2.4$ & $1.9\pm0.54$ \\
1126b & 0.88 & $16\pm2.9$ & $10\pm2.7$ & $2.1\pm0.81$ \\
0520a & 1.3 & $22\pm5.3$ & $14\pm4.7$ & $2.2\pm0.79$ \\
0320a & 1.5 & $21\pm3.8$ & $6.6\pm1.6$ & $1.7\pm0.33$ \\
\enddata
\tablenotetext{a}{The thermal pressure scale height, defined as
$H_P\equiv\int_{-\infty}^\infty P_{\rm therm}(z)dz/[2P_{\rm therm}(0)]$,
where $P_{\rm therm}(z)$ is the time-averaged vertical profile of gas plus
radiation pressure, in units of the fiducial scale height $H$ used in the
simulation grid.}
\tablenotetext{b}{The time-average of the thermal time, defined as the ratio
of instantaneous thermal energy content in the simulation domain to the
instantaneous horizontally averaged radiative flux emerging from the top and
bottom faces.  Errors indicate one standard deviation in the time average.}
\tablenotetext{c}{The time average of the ratio of box-averaged radiation
pressure $E/3$ to box-averaged gas pressure $p$.
Errors indicate one standard deviation in the fluctuations
of the ratio about the time average.}
\tablenotetext{d}{The time average of the ratio of maximum to minimum density
at the midplane $z=0$.  Errors indicate one standard deviation in the
fluctuations of the ratio about the time average.}
\end{deluxetable*}

\section{Energetics and Thermodynamics}
\label{sec:energetics}

The shearing box equations of motion as implemented in our code conserve
total energy to high accuracy \citep{hir06}.  Energy originates in
the work done by turbulent stresses on the shearing radial walls of the box
and ultimately escapes from the top and bottom of the box, largely in the
form of photons.  Long-term equilibrium requires that nearly all the work done
by the walls be dissipated and that all the heat generated be vented.
In this section we analyze this energy flow
and dissipation in detail as a function of height.

\subsection{Total Energy Conservation}
\label{sec:energy}

Equations (\ref{eq:cont})-(\ref{eq:induction}) and (\ref{eq:fsb})
together imply a total energy conservation equation
\begin{eqnarray}
{\partial\over\partial t}\left({\cal E}_{\rm mech}+{\cal E}_{\rm therm}\right)
&+&\nabla\cdot\left({\bf F}_{\rm mech}+
{\bf F}_{\rm therm}\right)\cr
&=&{\bf v}\cdot\left(\nabla\cdot{\sf P}+
{\bar{\kappa}^{\rm R}\rho\over c}{\bf F}\right).
\label{eq:energytot}
\end{eqnarray}
The grid scale numerical heating $\tilde{Q}$ has been lost, as our total energy
numerical scheme is designed to capture grid scale losses of magnetic and
kinetic energy and convert them into heat, which is promptly used to create
photons.  There are also numerical energy
sources and sinks that should be present on the right hand side of this
equation due to density and internal energy floors and a velocity
cap (see appendix of \citealt{hir06}), but these are negligible and we ignore
them here.

The energy density in equation (\ref{eq:energytot}) has two parts.  The first
is mechanical, being the sum of kinetic, effective gravitational potential,
and magnetic energy densities:
\begin{equation}
{\cal E}_{\rm mech}\equiv{1\over2}\rho v^2+\rho\phi+{B^2\over8\pi}.
\end{equation}
The second is thermal, being the sum of the gas and radiation internal
energy densities:
\begin{equation}
{\cal E}_{\rm therm}\equiv e+E.
\end{equation}
The energy flux vector also has two similar pieces.  The mechanical energy
flux is the sum of kinetic energy flux, effective gravitational potential
energy flux, flux of work done by artificial viscosity and by gas and radiation
pressures, and Poynting flux:
\begin{equation}
{\bf F}_{\rm mech}\equiv{1\over2}\rho v^2{\bf v}+\rho\phi{\bf v}+
{\sf q}\cdot{\bf v}+p{\bf v}+{\sf P}\cdot{\bf v}+
{c\over4\pi}{\bf E}\times{\bf B},
\label{eq:fmech}
\end{equation}
where ${\bf E}=-{\bf v}\times{\bf B}/c$ is the electric field in ideal
MHD.  The thermal energy flux is the sum of gas and radiation internal
energy advection and heat transport by radiative diffusion,
\begin{equation}
{\bf F}_{\rm therm}\equiv(e+E){\bf v}+{\bf F}.
\label{eq:ftherm}
\end{equation}

The right hand side of equation (\ref{eq:energytot}) is an artificial set
of energy source and sink terms that result from our use of flux-limited
diffusion to handle radiation transport.  These terms exactly cancel in
the optically thick limit.  They would also cancel to lowest order in $v/c$
if we were using the full radiation momentum equation (eq. [9] of
\citealt{sto92})
instead of the diffusion equation (\ref{eq:raddiff}).  These terms contribute
negligibly to the overall energy balance in our simulations, and we
ignore them from now on.

Assuming an equilibrium has been established over long time scales,
the energy conservation equation (\ref{eq:energytot}) can be rewritten
to show that the local divergence of the time and horizontal average of
the vertical energy flux is given by the local rate at which work is
being done on the fluid by the shearing walls at that height $z$, i.e.
\begin{equation}
{d\over dz}\left(\left<F_{{\rm mech},z}\right>+
\left<F_{{\rm therm},z}\right>\right)={3\over2}\Omega\left<\tau_{xy}(z)\right>,
\label{eq:stressflux}
\end{equation}
where angle brackets denote horizontal and time-averages at a particular
height $z$.  For example,
\begin{eqnarray}
\left<F_{{\rm mech},z}(z)\right>&=&\cr
{1\over\Delta tL_xL_y}&&\int_0^{\Delta t} dt
\int_{-L_x/2}^{L_x/2}dx\int_{-L_y/2}^{L_y/2}dy F_{{\rm mech},z},
\end{eqnarray}
where $\Delta t$ is the simulation duration minus the first 10 orbits when
the MRI was still in its growth phase.  All time-averages presented in this
paper use this time interval.

The quantity $\left<\tau_{xy}(z)\right>$ is the time-averaged $xy$ component of
the Reynolds and Maxwell stresses at height $z$, plus a negligibly small
contribution from radiation viscosity,
\begin{eqnarray}
\left<\tau_{xy}(z)\right>&\equiv&{1\over \Delta tL_y}\int_0^{\Delta t}dt
\int_{-L_y/2}^{L_y/2}dy\Biggl(\rho v_x\delta v_y\cr
&&-{B_xB_y\over4\pi}+P_{xy}\Biggr).
\end{eqnarray}
Here $\delta v_y\equiv v_y+3\Omega x/2$ is the perturbation of azimuthal
velocity from the average shear flow, and the $y$-integral is done either on
the inner or outer radial wall of the box \citep{haw95}.  When vertically
integrated, the relative contributions of Maxwell and Reynolds stresses
are remarkably constant from simulation to simulation: they are
85\% and 15\% respectively, to within 1\% for all six simulations.
The radiation viscosity contributes at most $10^{-4}$ of the total vertically
integrated stress.

\begin{figure}
\plotone{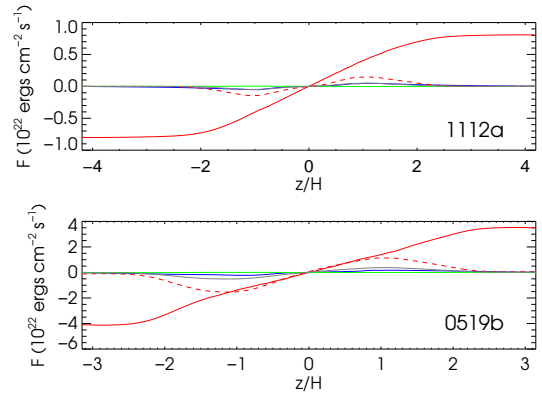}
\caption{Vertical profiles of the most important contributors to the
horizontally and time-averaged vertical energy flux in simulations 1112a
(top) and 0519b (bottom).  The different curves are diffusive radiation
energy flux (red), advected radiation energy flux (dashed red), flux of
radiation pressure work (gray), Poynting flux (blue), and advected gas
internal energy flux (green).  The radiation pressure work and Poynting flux
curves nearly coincide in simulation 1112a.}
\label{fig:energyflux}
\end{figure}

Figure \ref{fig:energyflux} shows the time-averaged vertical profiles of the
dominant contributions to the vertical thermal and mechanical energy fluxes
in simulations 1112a (one of the two lowest radiation to gas pressure ratio
simulations we consider in this paper) and 0519b (the highest radiation to
gas pressure ratio).  As in the case of gas pressure dominated simulations
\citep{hir06} and simulations with comparable gas and radiation pressure
\citep{kro07},
radiative diffusion is the dominant process of vertical energy transport
in simulation 1112a.  However, we now see that at the highest levels of
radiation pressure support simulated thus far (0519b), radiation advection
is just as important in the midplane regions.

\begin{figure*}
\plotone{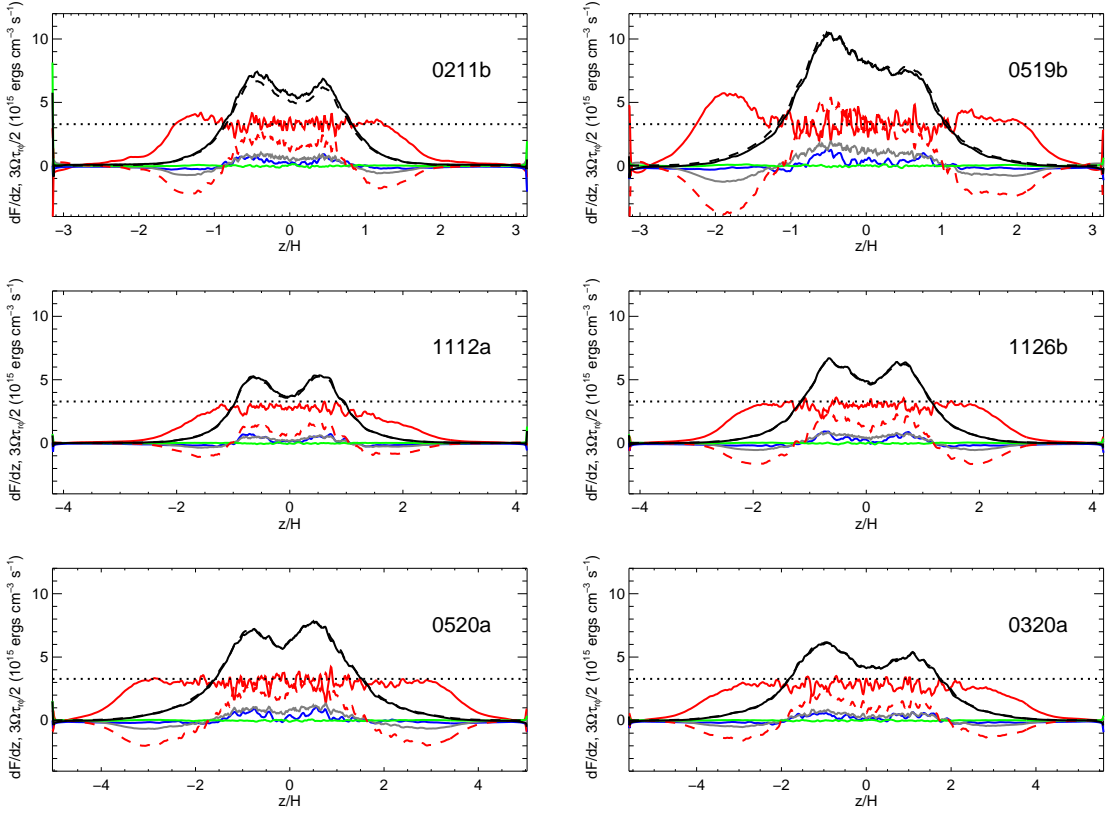}
\caption{Divergences of the horizontally-averaged and time-averaged vertical
thermal and mechanical energy fluxes.
The different colored curves show the divergences of various
vertical fluxes:  advected gas internal energy flux (green), advected
radiation energy flux (dashed red), diffusive radiation energy flux
(solid red), flux of radiation pressure work (gray), and Poynting flux (blue).
The total of these flux divergences is shown as the solid black curve, which
matches very well the profile of stress times rate of strain (dashed black
curve), in agreement with equation (\ref{eq:stressflux}).  (We have neglected
the flux of gas pressure work, which being 2/3 of the advected gas
internal energy flux is negligible.)  The horizontal dotted line indicates
the fiducial dissipation $c\Omega^2/\kappa_{\rm es}$ expected from hydrostatic
equilibrium in the radiation pressure dominated limit.
\label{fig:divenergyflux}}
\end{figure*}

Figure \ref{fig:divenergyflux} depicts the time-averaged vertical profiles
of the various contributions to the divergences of the vertical thermal
and mechanical energy fluxes, and compares their sum to the vertical profile of
stress times rate of strain.  Equation (\ref{eq:stressflux}) is accurately
satisfied by all the simulations.  The radiation diffusion flux divergence
approximately matches the $c\Omega^2/\kappa_{\rm es}$ value required by
hydrostatic equilibrium, presumably because departures from this equilibrium
would result in very fast readjustments on the dynamical time scale.
All the remaining flux divergence components are all positive in the
midplane regions and negative further out.  The energy injected by the stresses
on the walls in the midplane regions is significantly larger than can be
carried away by radiative diffusion and still maintain vertical hydrostatic
equilibrium.  This excess energy is therefore transported outward
by the other forms of energy flux, among which radiation advection is the
most important.  {\it This process is generally completely ignored
in standard accretion disk models.}

\subsection{The First Law of Thermodynamics: $PdV$ Work and Radiative Damping}
\label{sec:thermodynamics}

By combining equations (\ref{eq:gasenergy}) and (\ref{eq:radenergy}), energy
conservation may also be described in terms of the first law of thermodynamics,
\begin{equation}
{\partial{\cal E}_{\rm therm}\over\partial t}+
\nabla\cdot{\bf F}_{\rm therm}=Q-p\nabla\cdot{\bf v}-
{\sf P}:\nabla{\bf v}.
\label{eq:firstlawthd}
\end{equation}
Here $Q\equiv\tilde{Q}-{\sf q}:\nabla{\bf v}$ is the local dissipation
rate per unit volume, being a sum of grid-scale conversions of magnetic
and kinetic energy into heat as well as energy dissipated by artificial
viscosity.  (The artificial viscosity dissipation is approximately 3/4 of
the grid-scale loss rate of kinetic energy in all of our simulations, and
has a similar time- and horizontally-averaged vertical profile.)
Again assuming an equilibrium has been established over long time scales,
equation (\ref{eq:firstlawthd}) implies that
\begin{equation}
{d\over dz}\left<F_{{\rm therm},z}\right>=
\left<Q-p\nabla\cdot{\bf v}-{\sf P}:\nabla{\bf v}\right>.
\label{eq:dissflux}
\end{equation}
In other words, the local divergence of the average vertical
thermal energy flux is equal to the local dissipation per unit volume
plus the rate at which gas and radiation pressure work is being done
on the plasma per unit volume.

\begin{figure*}
\plotone{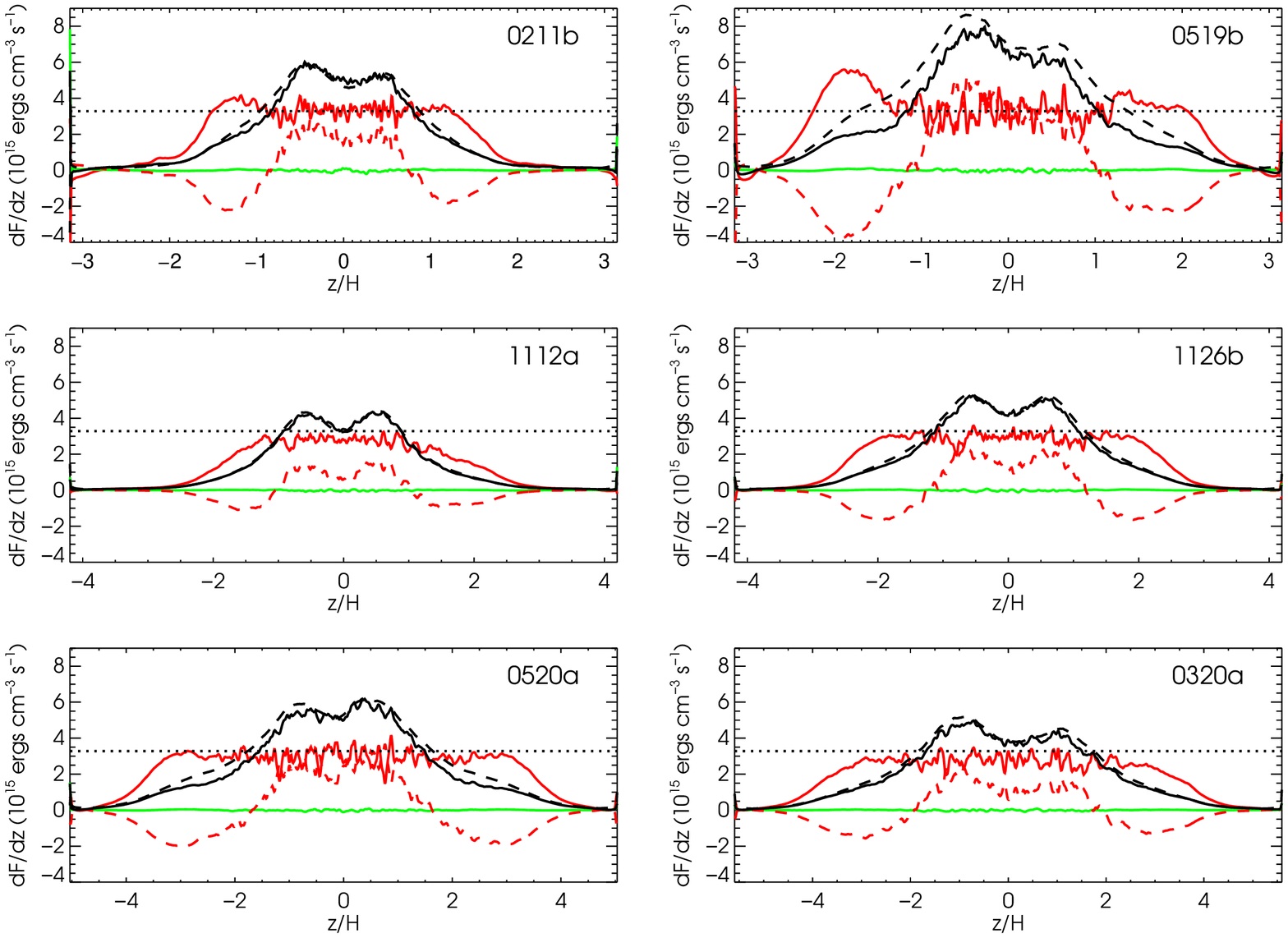}
\caption{Divergences of the horizontally-averaged and time-averaged vertical
thermal energy fluxes.  The green curve shows the divergence of
the advected gas internal energy flux, the dashed red curve shows the divergence
of the advected radiation energy flux, the solid red curve shows the divergence
of the diffusive radiation energy flux, and the solid black curve shows
the total of these three curves, i.e. $\left<dF_{{\rm therm},z}/dz\right>$.
The dashed black curve shows the horizontal and time-averaged dissipation
rate per unit volume $<Q>$.  This exceeds the divergence of the thermal flux,
the most notable discrepancy being in simulation 0519b, which has the highest
average radiation to gas pressure ratio.
The horizontal dotted line indicates
the fiducial dissipation $c\Omega^2/\kappa_{\rm es}$ expected from hydrostatic
equilibrium in the radiation pressure dominated limit.
\label{fig:divthermalflux}}
\end{figure*}

Equation~(\ref{eq:dissflux}) appears to be familiar, but its form hides
some subtleties peculiar to radiation dominated disks.  These new effects
emerge when comparing the two sides of this equation.
Figure \ref{fig:divthermalflux} depicts the vertical profiles of the advection
and radiative diffusion contributions to the divergence of the vertical thermal
flux, as well as the numerical dissipation profile, for all six of our
simulations.  Because our numerical dissipation $Q$ does not include
damping by radiative diffusion, one might have expected it to be smaller
than the divergence of the thermal flux.  In fact the opposite is true, with
the largest discrepancy occurring in the most radiation pressure dominated
simulation 0519b.  Moreover, the absence of an explicit place for radiative
damping suggests that something is missing because that process should
contribute to the total dissipation rate.

\begin{figure*}
\plotone{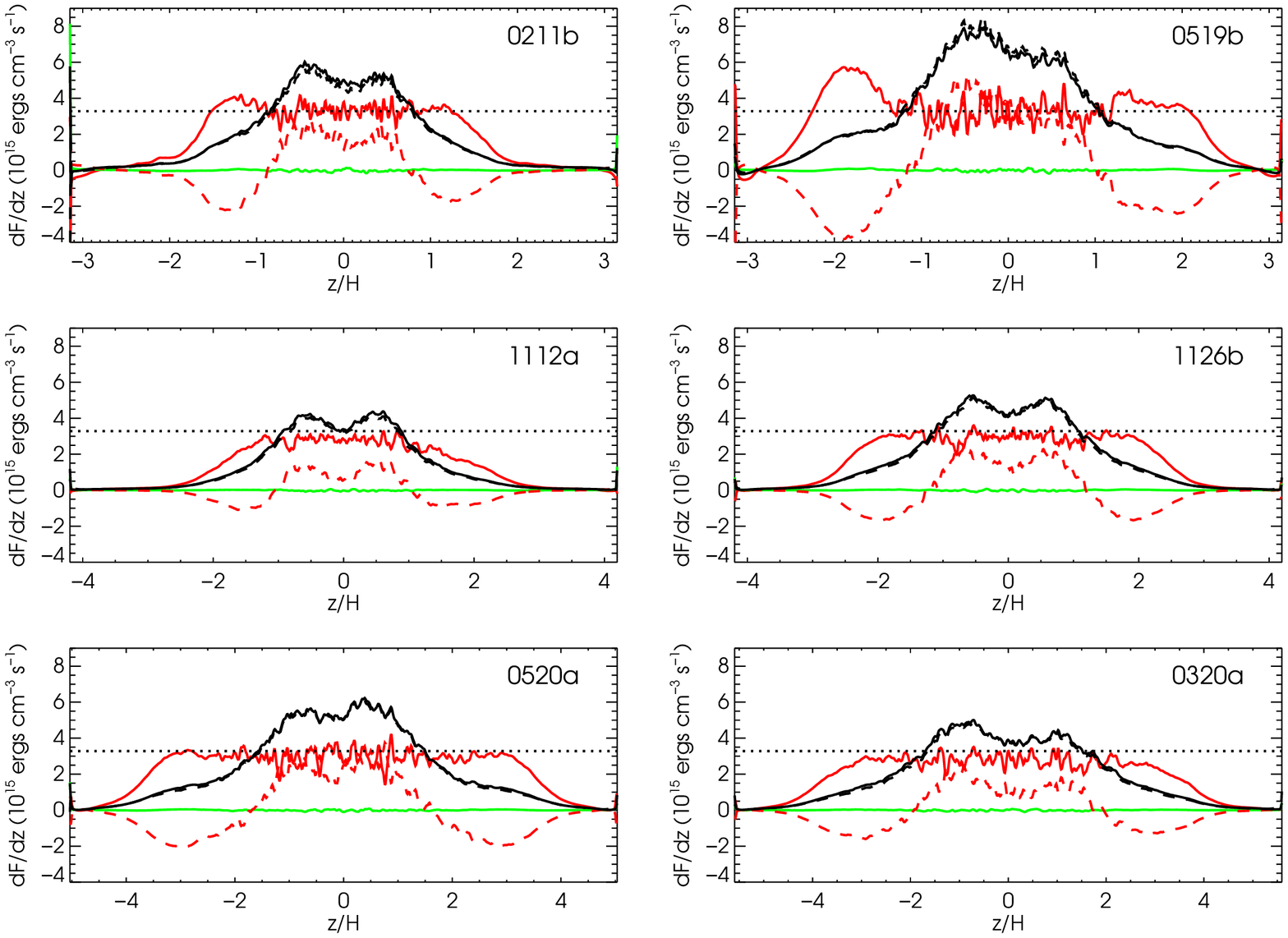}
\caption{Same as Figure~\ref{fig:divthermalflux} except now the dashed
black curve shows the horizontal and time-averaged dissipation plus the
rate at which pressure work is being done on the plasma per unit volume,
i.e. $<Q-{\sf P}:\nabla{\bf v}>$.  (The gas pressure work
$-p\nabla\cdot{\bf v}$ is negligible, and has been neglected.)  This agrees
very well with the divergence of the thermal flux, in agreement with equation
(\ref{eq:dissflux}).  The horizontal dotted line indicates
the fiducial dissipation $c\Omega^2/\kappa_{\rm es}$ expected from hydrostatic
equilibrium in the radiation pressure dominated limit.
\label{fig:divthermalfluxsilk}}
\end{figure*}

Figure \ref{fig:divthermalfluxsilk} shows the same vertical
profiles of thermal flux divergence, but a dissipation rate corrected
by the $PdV$ work.  With this adjustment, the time-averaged first law
of thermodynamics (\ref{eq:dissflux}) is accurately satisfied in each of the
simulations.  From this fact, we reach two important conclusions.  First,
there is a significant conceptual flaw in
classical time-steady accretion disk models.  These models assume that the
divergence of the diffusive radiation flux completely defines the left hand
side of the first law of thermodynamics, while the local dissipation rate
is the only contribution to its right hand side.  But we have just seen that
in radiation dominated disks {\it both} of these simplifications are wrong:
radiation advection must be included with radiative diffusion, and the $PdV$
work terms are important.
Second, radiative damping is actually {\it included} in the $PdV$ work terms.
In our scheme, the dissipation associated with it is conveyed by the
diffusion equation, but the energy it dissipates flows into the gas
via the pressure work terms in the first law of thermodynamics.

It is perhaps helpful to make more explicit the actual dissipation associated
with radiative diffusion.  The first law of thermodynamics
(\ref{eq:firstlawthd})
can be combined with the radiative diffusion equation (\ref{eq:raddiff})
to derive an equation for the evolution of the entropy per unit mass $s$ of
the gas plus radiation mixture.  Restricting consideration to optically thick
regions for simplicity, this equation is
\begin{equation}
\rho\left({\partial s\over\partial t}+{\bf v}\cdot\nabla s\right)+
\nabla\cdot\left({{\bf F}\over T}\right)={4acT\over3\bar{\kappa}^{\rm R}\rho}
\left(\nabla T\right)^2+{Q\over T}.
\label{eq:entropyevolution}
\end{equation}
The last term on the left hand side is the divergence of the entropy flux
due to radiative diffusion.  The first and second terms on the right hand
side are sources of entropy (dissipation) due to radiative diffusion and
grid scale losses plus artificial viscosity, respectively.  Assuming an
equilibrium has been established on long time scales, this equation becomes
\begin{equation}
<\rho {\bf v}\cdot\nabla s>+{d\over dz}\left<{F_z\over T}\right>=
\left<{4acT\over3\bar{\kappa}^{\rm R}\rho}\left(\nabla T\right)^2\right>+
\left<{Q\over T}\right>.
\label{eq:entropy}
\end{equation}
Advective and diffusive transport of entropy is therefore balanced by
dissipation.  The dissipation associated with radiative damping of
fluctuations in the turbulence is in the first term on the right hand side, but
this term also includes entropy generation due merely to the photon diffusion
down the background average vertical temperature gradient.  An analogous
situation holds in stars with radiative envelopes.  Under static conditions,
the outward luminosity at every radius in the envelope is constant.  Because
the temperature at the base of the envelope is higher than the temperature
of the photosphere, the entropy leaving the photosphere exceeds the entropy
entering the base of the envelope.  The source of this entropy increase is the
dissipative nature of photon diffusion itself, and is
not associated with dissipative release of energy that must then be
transported away to establish thermal equilibrium.

Because of this non-energy releasing dissipation, we cannot use equation
(\ref{eq:entropy}) to calculate the true dissipative heating due to radiative
damping of fluctuations.  Instead, we must try and use the pressure
work terms in the first law of thermodynamics (\ref{eq:dissflux}).  As
we noted above, it is through these terms that energy flows into the gas from
turbulent fluctuations, and ultimately dissipates by radiative diffusion.

However, somewhat surprisingly, the time-averaged pressure work corrections
$-<p\nabla\cdot{\bf v}>-<{\sf P}:\nabla{\bf v}>$
are typically {\it negative}, in contrast
to the case of shearing boxes without vertical gravity studied by
\citet{tur02,tur03}.  This sign change indicates that the plasma does
net work through expansion even though these corrections must include
radiative damping, which would be positive (it is dissipative).  The fact
that the overall pressure work corrections are negative means that the
gas loses more thermal energy by driving expansion away from the midplane
than it gains by radiative damping.

Fortunately, these two contributions to the pressure work corrections
(pumping expansion and radiative damping) can be separated analytically.
Considering only the optically thick regions, write the thermal pressure
$P_{\rm therm}=p+E/3$ and fluid velocity as
$P_{\rm therm,av}+\delta P_{\rm therm}$
and ${\bf v}_{\rm av}+\delta{\bf v}$, respectively.  Here the ``av'' subscript
denotes horizontal average and $\delta P_{\rm therm}$ and $\delta{\bf v}$
therefore have zero horizontal average by definition.  The divergence of
$\delta{\bf v}$ also
has zero horizontal average, as one can show by integrating by parts and using
the shearing box boundary conditions.  Hence the horizontal average of the
thermal pressure work done on the plasma under optically thick conditions is
\begin{eqnarray}
{1\over L_xL_y}\int_{-L_x/2}^{L_x/2} dx\int_{-L_y/2}^{L_y/2}dy
\left(-p\nabla\cdot{\bf v}-{\sf P}:\nabla{\bf v}\right)=\cr
-P_{\rm therm,av}{\partial\over\partial z}v_{{\rm av},z}(z,t)\cr
+{1\over L_xL_y}\int_{-L_x/2}^{L_x/2} dx\int_{-L_y/2}^{L_y/2}dy
(-\delta P_{\rm therm}\nabla\cdot\delta{\bf v}).
\label{eq:pressureworksplit}
\end{eqnarray}
After time-averaging, the first term on the right hand side represents pumping
of vertical mechanical motions.  This term will be negative if the plasma
undergoes vertical expansion in a horizontally-averaged sense, i.e. the plasma
will do net work.  We will henceforth call this term the mechanical pumping
term.

The second term on the right hand side of equation (\ref{eq:pressureworksplit})
is the radiative damping.  For a sinusoidal adiabatic fluctuation, this term
would vanish identically in the time average, because $\delta P_{\rm therm}$
and $\nabla\cdot\delta{\bf v}$ are 90 degrees out of phase.  Radiative
diffusion removes this cancellation by making $\delta P_{\rm therm}$ lead
$\nabla\cdot\delta{\bf v}$ by a little more than 90 degrees.  For example,
if the plasma is locally compressed nearly adiabatically, the temperature
rises and radiative diffusion to the cooler surrounding regions increases.
This diffusion is fastest as the point
of maximum compression is approached, while at the same time the rate of
compression is slowing down.  The temperature therefore starts to drop
just before the point of maximum compression, i.e. the maximum in temperature
is reached before the point of maximum compression.
The same holds on the expansion cycle, where the minimum in temperature
is reached before the point of maximum expansion.  Hence there is a greater
than 90 degree lead in $\delta P_{\rm therm}$ relative to
$\nabla\cdot\delta{\bf v}$.\footnote{Very short wavelength fluctuations are
approximately isothermal, not adiabatic, because radiative diffusion is so
rapid on short length scales.  Finite (as opposed to infinitely rapid)
radiative diffusion in this limit also produces the same greater than 90
degree lead between thermal pressure and $\nabla\cdot\delta{\bf v}$.  In a
compression phase in this case, the work being done on the plasma causes it
to be a little hotter than it would be if it were isothermal, resulting
in a slightly greater thermal pressure. But this excess pressure must then
drop as the point of maximum compression is reached because radiative diffusion
is most rapid there, returning the pressure to the isothermal value.
Hence the maximum in thermal pressure leads the point of maximum compression.}
This phase offset results in a net positive time average of
$-\delta P_{\rm therm}\nabla\cdot\delta{\bf v}$.  The plasma has net
positive work done on it, and that excess work is dissipated by radiative
diffusion.
In principle, the mechanical pumping term of equation
(\ref{eq:pressureworksplit}) may still include a small positive contribution
from the radiative damping of purely vertical acoustic
waves, but we show in section~\ref{sec:verticalwaves} below that this
contribution is negligible.

\begin{figure}
\plotone{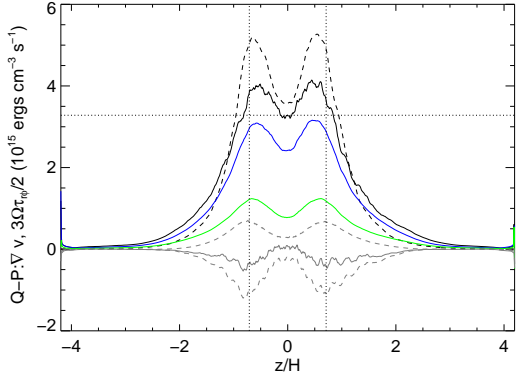}
\caption{Time and horizontally-averaged vertical profiles of stress times
rate of strain $3\Omega\tau_{r\phi}/2$ (dashed), grid scale magnetic energy
dissipation (blue), grid scale kinetic energy dissipation plus artificial
viscosity dissipation (green), and $-{\sf P}:\nabla{\bf v}$ work (grey), for
simulation 1112a. The total of these last three, i.e.
$Q-{\sf P}:\nabla{\bf v}$, is shown as the solid black curve and matches
the time and horizontally-averaged profile of thermal energy flux divergence.
The lower gray dashed curve shows the time-averaged profile of
$-(E_{\rm av}/3)dv_{z,{\rm av}}/dz$, i.e. minus the horizontally
averaged radiation pressure times the vertical derivative of the horizontally
averaged vertical velocity.  This represents the spatial profile of work done
to pump vertical mechanical motions.  The difference between this and the total
pressure work profile is given by the upper dashed gray curve, which represents
the radiative damping contribution to the dissipation.  The horizontal
dotted line again indicates the fiducial dissipation $c\Omega^2/\kappa_{\rm es}$
for a radiation pressure dominated hydrostatic equilibrium.  Vertical dotted
lines indicate one pressure scale height $H_P$ away from the midplane.
\label{fig:dissstress1112a}}
\end{figure}

\begin{figure}
\plotone{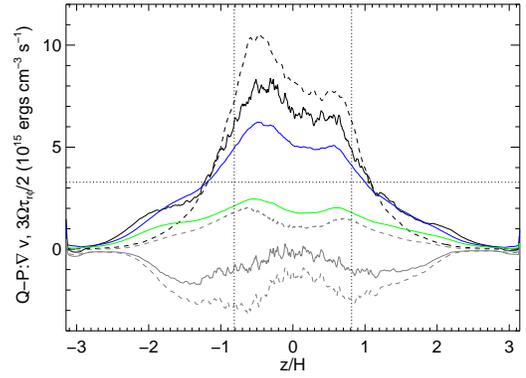}
\caption{Same as Figure \ref{fig:dissstress1112a}, except for simulation
0519b.
\label{fig:dissstress0519b}}
\end{figure}

Figures \ref{fig:dissstress1112a} and \ref{fig:dissstress0519b} depict
the time-averaged vertical profiles of stress times rate of strain and
the different contributions to dissipation and mechanical work for simulations
1112a and 0519b.  We did not save high time resolution data over the
entire simulation duration to enable
us to directly compute the last term on the right hand side of equation
(\ref{eq:pressureworksplit}), so the radiative damping profiles in these figures
were computed from subtracting the vertical profile of the first term from
the vertical profile of the left hand side.  We also neglected gas pressure
work here as it is very small in these simulations.

Two features are worth noting
about these profiles.  First, even after time-averaging, there remain spatial
fluctuations in the total mechanical work profile as well as the profile of
mechanical pumping.  Those fluctuations are completely absent in the radiative
damping profile, which is as smooth as the other (magnetic and kinetic)
dissipation profiles.  The radiative damping profile is also very similar in
shape to these other dissipation profiles.  Second, when compared to the
numerical dissipation, the radiative damping and mechanical pumping,
as well as the net total pressure work, are clearly relatively more important
in 0519b (the simulation with the highest radiation to gas pressure ratio)
than in 1112a.  As much as 22 percent of the work done by the shearing walls
ends up being dissipated by radiative diffusion in simulation
0519b.\footnote{The radiative
damping percentages of the work done by the shearing walls are 12 and
15 percent for simulations 1112a and 1126b, respectively.  These are much
higher than the 1.3 and 0.7 percent values that we stated in \citet{hir09}
(see end of section 3 in that paper).  Those previous numbers came from
integrating up the total pressure work near the midplane where it is (barely)
positive (see the solid gray curve of Figure~\ref{fig:dissstress1112a}),
indicating net damping.  Our new analysis, which cleanly separates
radiative damping from work associated with vertical expansion, shows that the
true radiative damping is much larger in these simulations and
peaks off the midplane (upper gray dashed curve of
Figure~\ref{fig:dissstress1112a}).}

\begin{figure}
\plotone{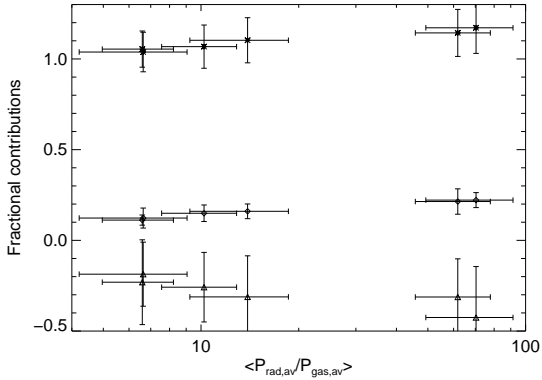}
\caption{Time-averaged ratios of various box-averaged contributions to
the dissipation and pressure work to the vertically integrated
stress times rate of strain, as a function of the average radiation to
gas pressure ratio of each simulation.  The upper set of points (crosses)
shows the fractional contribution of the numerical dissipation $Q$
(grid scale losses of magnetic and kinetic energy as well as artificial
viscosity).  The middle set of points (diamonds) shows the radiative damping,
and the bottom set of points (triangles) shows the radiation pressure work
associated with pumping of vertical mechanical motions.  Error
bars indicate one standard deviation in the time-averages.}
\label{fig:disscontributions}
\end{figure}

\begin{figure}
\plotone{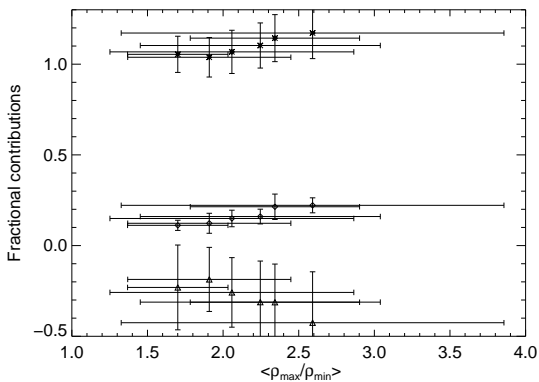}
\caption{Same as Figure \ref{fig:disscontributions}, only now as a
function of the time-averaged density contrast at the midplane of each
simulation.}
\label{fig:disscontributionsrho}
\end{figure}

Figure \ref{fig:disscontributions} illustrates this trend of increasing
relative importance of the pressure work corrections with growing radiation
to gas pressure ratio.  A larger rate of radiative damping presumably requires
larger density fluctuations; to test this supposition, we have
measured the time average of the ratio of maximum to minimum density at
the midplane in each of the simulations, and the results are listed in
Table~\ref{simprops}.  Density fluctuations in the
midplane regions do indeed become slightly larger with increasing radiation
to gas pressure.  Figure~\ref{fig:disscontributionsrho} shows the fractional
pressure work as a function of the time-averaged density contrast at the
midplane.  The large horizontal error bars reflect the large variations in
the density contrast, but the radiative damping points of this figure are
clearly consistent with the trend observed in non-stratified
shearing boxes by \citet{tur03} (see top panel of their Figure 7).

\begin{figure}
\plotone{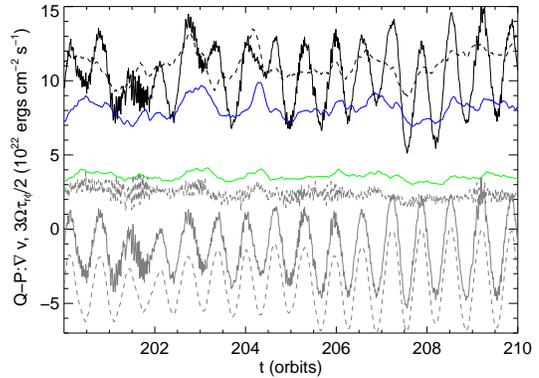}
\caption{Time dependence over a ten orbit period of the vertically integrated
stress times rate of strain and various contributions to the dissipation
and work shown in Figure~\ref{fig:dissstress0519b} for simulation 0519b.  The
colors and line styles correspond to the same quantities as in
Figures~\ref{fig:dissstress1112a} and \ref{fig:dissstress0519b}.
\label{fig:dissstressvst0519b}}
\end{figure}

There is considerable structure in the temporal behavior of the pressure work
terms.  A ten orbit segment of the time-dependence of the vertical integral of
the stress times rate of strain, the dissipation terms, and the pressure
work terms for simulation 0519b is shown in
Figure~\ref{fig:dissstressvst0519b}.  The mechanical pumping work (lower
gray dashed curve) shows clear oscillatory behavior on time scales of
order the orbital period.  This is completely absent in the radiative damping
(upper gray dashed curve), which instead clearly exhibits much higher
frequency variability.  Both also exhibit much longer time scale variation.

\begin{figure}
\plotone{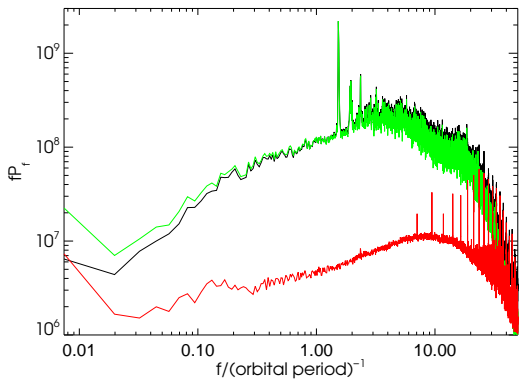}
\caption{Vertically-integrated power spectra of the $-{\sf P}:\nabla{\bf v}$
work (black), the mechanical pumping portion of this work (green), and the
radiative damping portion of this work (red), for simulation 0519b.}
\label{fig:mpdvpowspec}
\end{figure}

Some aspects of this behavior can be immediately understood by
Fourier transforming the time dependence and plotting the temporal power
spectrum of the vertically integrated mechanical pumping and radiative damping
terms. The result is shown in Figure~\ref{fig:mpdvpowspec}.
The large oscillations seen in the mechanical pumping in
Figure~\ref{fig:dissstressvst0519b} are reflected in a series of discrete
sharp peaks.  These peaks represent standing vertical acoustic waves that
are trapped in the box.  Despite their prominence, we show below in
section~\ref{sec:verticalwaves} that they actually contribute negligibly
to the energetics of the disk.  The radiative damping power spectrum exhibits
much higher frequency power, including
a set of high frequency, extremely sharp spikes.  This is
consistent with the high frequency variability seen in the time domain of
Figure~\ref{fig:dissstressvst0519b}.  To better understand this variability,
we must first examine the types of fluctuation that can produce radiative
damping, to which we now turn.

\section{Varieties of Radiation Pressure Fluctuations}
\label{sec:pradmodes}

It is radiation pressure (i.e. temperature) fluctuations that give rise
to photon diffusion and therefore radiative damping.  In this section we discuss
two distinct types of such fluctuations that clearly play a role in our
simulations.  We will also see in section~\ref{sec:verticaladvection} that
understanding both radiation advection and energy transport by Poynting flux
will likewise be aided by a prior understanding of radiation pressure fluctuations.

In classical MHD theory (in which radiation pressure is negligible), linear
compressible waves are classified into ``fast" and ``slow" modes.  One way to
understand qualitatively this distinction is to note that the magnetic and gas
pressure perturbations are exactly in phase in the former mode and exactly
out of phase in the latter; it is the partial cancellation of the pressure
perturbation in the slow mode that causes its low propagation speed.
Our simulations do not exactly correspond to this categorization in two ways:
the fluctuations are often nonlinear; and radiation pressure is both significant
and only imperfectly coupled to the fluctuations (that is what radiative damping
is all about, of course).  Nonetheless, this conceptual division remains
useful because large total (gas plus radiation plus magnetic) pressure
fluctuations at a given length scale tend to have much higher frequency
than those with small total pressure fluctuations.  The large total pressure
fluctuations therefore propagate on an approximately stationary background
set by the slowly evolving small total pressure fluctuations.  We will call
the former ``acoustic waves" and the latter ``isobaric waves" in order to
stress this distinction.

\subsection{Acoustic waves}

\begin{figure*}
\plotone{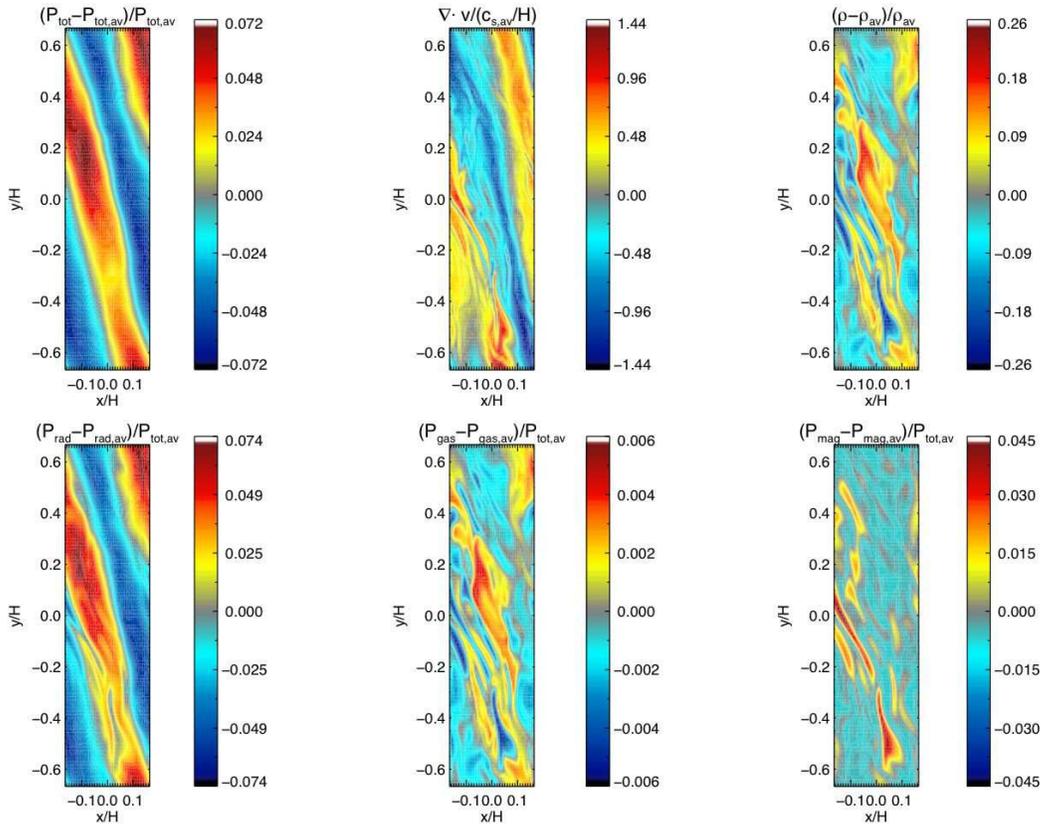}
\caption{Spatial distribution of various quantities in the $z=0$ midplane
of simulation 0519b at epoch 200.3 orbits (near the time of peak amplitude
of the $n_x=9$, $n_y=1$ nonaxisymmetric wave shown in
Figure~\ref{fig:ptotwaveamplitudes}).  On top from left to right are the
total (radiation plus gas plus magnetic) pressure perturbation, scaled
with the horizontally averaged total pressure at the midplane; the fluid
velocity divergence, scaled with the horizontally averaged midplane sound
speed divided by the fiducial scale height; and the density perturbation
scaled with the horizontally averaged midplane density.  On the bottom
from left to right are the radiation pressure perturbation, gas pressure
perturbation, and magnetic pressure perturbation, respectively, scaled
with the horizontally averaged total pressure at the midplane.
\label{fig:spiralwavesmidplaneplot}}
\end{figure*}

\begin{figure*}
\plotone{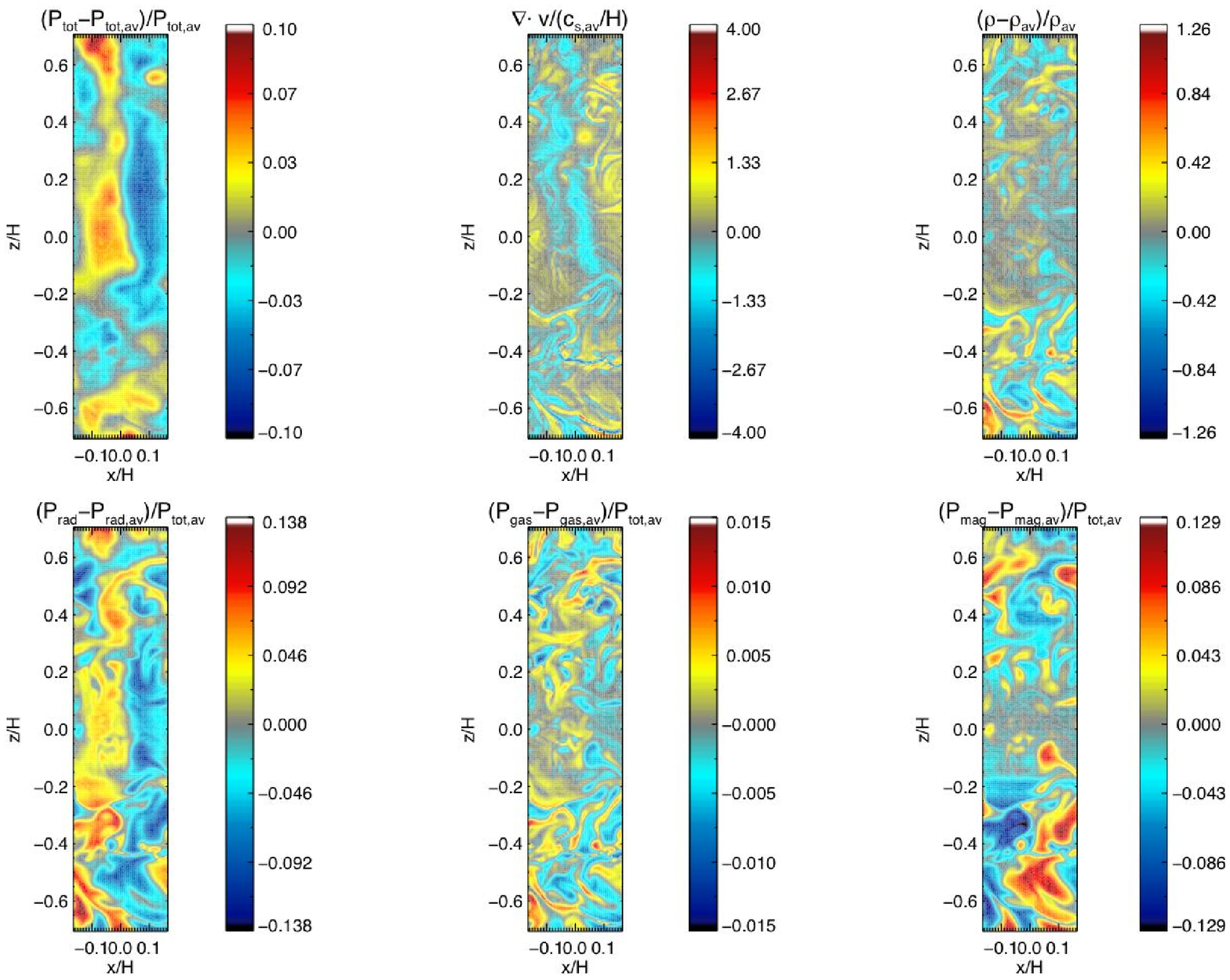}
\caption{Spatial distribution of various quantities in the $y=0$ vertical
plane of simulation 0519b at epoch 200.3 orbits, depicting only the
regions within $\pm0.7H$ of the $z=0$ midplane.  The quantities shown
are the same as in Figure~\ref{fig:spiralwavesmidplaneplot}.
The perturbations are all measured with respect to, and scaled by, locally
horizontally averaged quantities at each height $z$.
\label{fig:spiralwavesxzplot}}
\end{figure*}

\citet{ago98} suggested that acoustic waves (i.e., fast magnetosonic waves)
would be the dominant contributor to radiative damping in radiation dominated
accretion disks.  We have already seen that standing vertical acoustic
waves are excited in the box.  In addition, strong nonaxisymmetric acoustic
waves are pervasive in all our simulations.  These waves are almost certainly
stochastically excited by the MRI turbulence itself \citep{hei09a,hei09b}.
In these waves, gas density, radiation pressure, and magnetic pressure
fluctuations all oscillate in phase.
Figure~\ref{fig:spiralwavesmidplaneplot}
depicts a snapshot of various quantities at the $z=0$ midplane of simulation
0519b at 200.3 orbits.  A wave pattern
is clearly evident in most of the fluid quantities shown, though it is cleanest
in the total pressure (upper left).  Figure~\ref{fig:spiralwavesxzplot} depicts
a vertical slice of the midplane regions at the same time.  Particularly
in the total pressure (upper left), the wave is once again evident in the
vertical yellow and blue stripes in the midplane region
$|z|\lesssim0.2-0.3H$
and primarily propagates in the horizontal direction.

\begin{figure}
\plotone{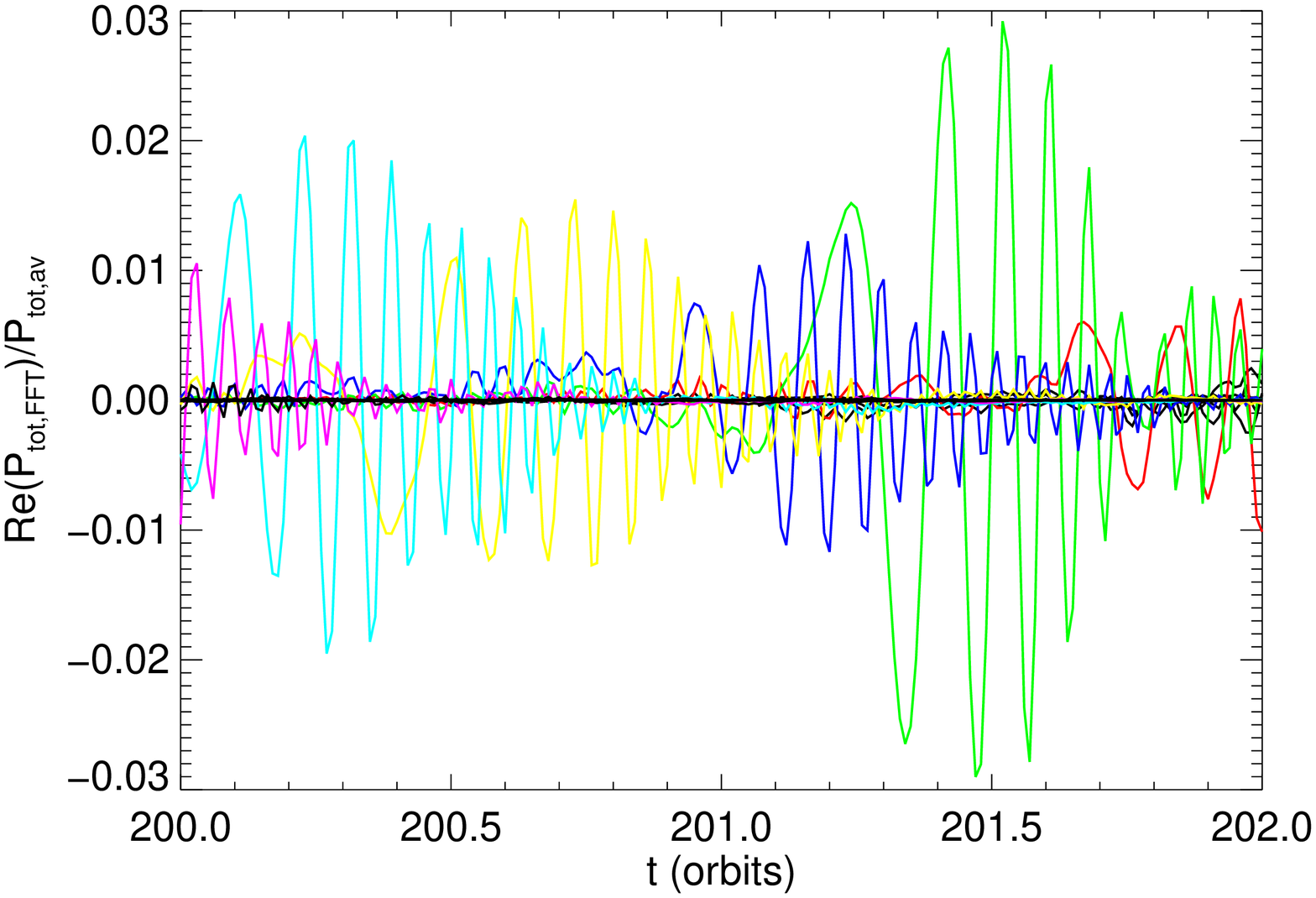}
\caption{Time dependence of the real part of the $n_y=1$ spatial Fourier
amplitude in total pressure at the $z=0$ midplane, scaled with the horizontally
averaged total pressure there.  Different colored curves correspond to
different values of $n_x$: $n_x=5$ (red), $n_x=6$ (green), $n_x=7$ (blue),
$n_x=8$ (yellow), $n_x=9$ (cyan), and $n_x=10$ (magenta).  Each of these
$n_x$ values correspond to wave vectors that happen to be swinging from
leading to trailing in or near the particular time interval shown.
Other values of $n_x$ are plotted as black curves.
\label{fig:ptotwaveamplitudes}}
\end{figure}

To better understand these nonaxisymmetric waves, it is helpful to project
the spatial variation of the $z=0$ midplane fluid quantities
onto the natural set of basis vectors $\exp[i(k_x(t)x+k_yy)]$ of the shearing
box, where
\begin{equation}
k_x(t)={2\pi n_x\over L_x}+{3\over2}\Omega k_y t,
\end{equation}
\begin{equation}
k_y={2\pi n_y\over L_y},
\end{equation}
and $n_x$ and $n_y$ are integer quantum numbers \citep{haw95,hei09b}.  The
largest azimuthal wavelength ($|n_y|=1$) waves always have the most power.
Figure~\ref{fig:ptotwaveamplitudes} shows the real part of the total pressure
Fourier amplitude as a function of time for $n_y=1$ and various values of
$n_x$.  In agreement with \citet{hei09b}, waves with different values of the
$n_x$ quantum number reach high amplitude at distinct times (in declining
$n_x$ order) when they
swing from leading to trailing, resulting in a series of wave pulses.  The
separation in time between these pulses is
determined entirely by the time interval between successive epochs at which
the shearing radial boundaries become exactly periodic in the radial
direction.  This shear time of the box is $\delta T_s=2L_y/(3\Omega L_x)$,
i.e.  $3\pi/4\simeq2.36$ inverse orbital periods.  We purposely chose the
200.3 orbits epoch for Figures~\ref{fig:spiralwavesmidplaneplot} and
\ref{fig:spiralwavesxzplot} as it corresponds to the time of peak amplitude
for one of these wave pulses ($n_x=9$, $n_y=1$).

\subsection{Isobaric waves}

The nonaxisymmetric acoustic wave pattern is evident in the spatial distribution
of density, shown in the top right panel of
Figure~\ref{fig:spiralwavesmidplaneplot}, but this pattern is markedly
perturbed by shorter length scale fluctuations.  Among the most prominant
are rarefied regions, e.g. near ($x=0.05$, $y=-0.5$), that are correlated with
regions of low radiation
pressure, low gas pressure, and high magnetic pressure (bottom left to right
panels of Figure~\ref{fig:spiralwavesmidplaneplot}, respectively).

These are a second kind of radiation pressure fluctuation, one in which
the magnetic pressure oscillates with very nearly the same amplitude as the
sum of gas and radiation pressure, but with opposite phase.  As a result,
the total pressure hardly changes.  This near-cancellation of pressure
fluctuations is characteristic of slow magnetosonic modes, in which
the cancellation is exact to $\sim O(\beta^{-1})$ when the plasma
$\beta \gg 1$.
When such modes are placed in a rotating shear flow whose rotation
rate declines outward, they become the magnetorotational instability.
We might naturally expect to see many such features in our shearing box
simulations, and some can be expected to grow to nonlinear amplitude.

\begin{figure*}
\plotone{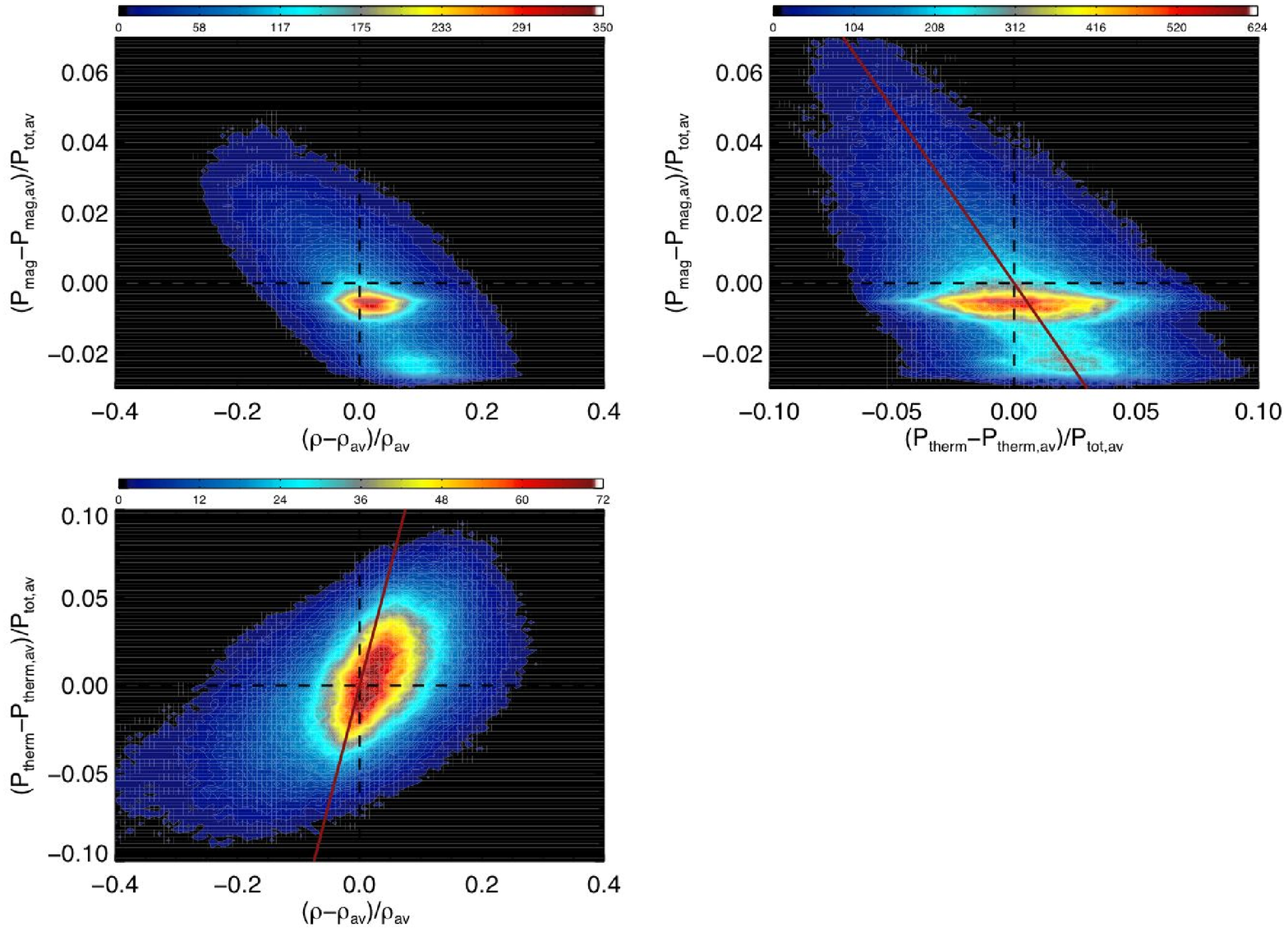}
\caption{Two dimensional distributions of the incidence of given values of
magnetic pressure fluctuations, density fluctuations, and thermal pressure
fluctuations in the $z=0$ midplane, averaged over the 200 to 202 orbit time
period in simulation 0519b.  Each distribution is normalized such that the two
dimensional integral over the perturbation variables is unity.  The red line
in the upper right plot shows the expected relation for isobaric
fluctuations, i.e. where the sum of the magnetic and thermal pressure
perturbations vanish.  The red line in the lower left plot shows the
expected relation for adiabatic perturbations:
$(P_{\rm therm}-P_{\rm therm,av})/P_{\rm therm,av}=\Gamma_1
(\rho-\rho_{\rm av})/\rho_{\rm av}$.
\label{fig:dpmagvsdpthermvsdrho}}
\end{figure*}

Figure~\ref{fig:dpmagvsdpthermvsdrho} shows correlation plots between
magnetic pressure fluctuations, density fluctuations, and thermal (i.e.
gas plus radiation) pressure fluctuations in the $z=0$ midplane of simulation
0519b near the 200 orbit epoch.  Most cells (i.e. most of the area as indicated
by the red and yellow regions in all three panels) house weakly negative
magnetic fluctuations, and the thermal pressure and density fluctuations in
these cells approximately obey the expected adiabatic relation for acoustic
waves: $\delta P_{\rm therm}/P_{\rm therm}=\Gamma_1\delta\rho/\rho$,
where $\Gamma_1\simeq4/3$ is the first generalized adiabatic index for
a gas and radiation mixture \citep{cha67}.  However, in the upper right plot
of magnetic pressure perturbation vs. thermal pressure perturbation, these
acoustic waves oscillate horizontally back and forth about a mean thermal
pressure perturbation that is set by a background of isobaric fluctuations.
This is indicated by the fact that the contours of cell-counts
stretch diagonally across this plot, showing that at large amplitude the
magnetic pressure and the total thermal (mostly radiation) pressure are
{\it anti-correlated}.  In addition, the {\it magnitudes} of the two
fluctuations are similar, so that $|\delta P_{\rm mag} + \delta P_{\rm therm}|$
is in general considerably smaller than
$|\delta P_{\rm mag}| + |\delta P_{\rm therm}|$.  The isobaric
modes also exhibit a clear anti-correlation between magnetic pressure and
density, presumably because the density is positively correlated with
the gas pressure component of the thermal pressure.

\begin{figure}
\plotone{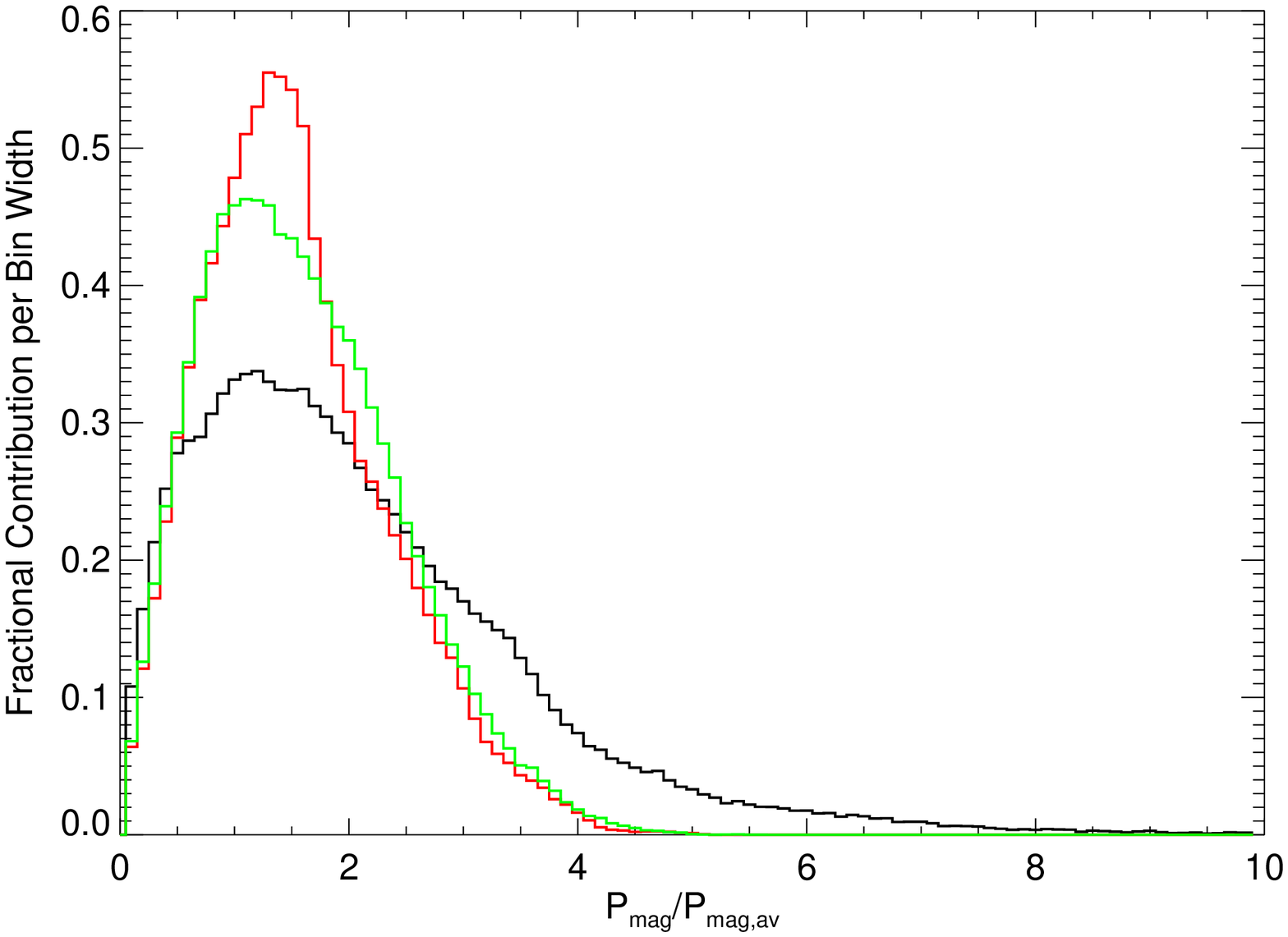}
\caption{The differential distribution of fractional contributions to the
total magnetic energy at a given height, as a function of local magnetic
pressure scaled by the average magnetic pressure at that height.  The
distributions are averaged from 200 to 202 orbits in simulation 0519b, and are
plotted for the midplane ($z=0$, black), $z=-0.5H$ (red), and $z=+0.5H$
(green).  Each of the distributions is normalized such that the integral
with respect to scaled magnetic pressure is unity.
\label{fig:pmagdist}}
\end{figure}

While most of the volume is occupied by weak magnetic field,
Figure~\ref{fig:pmagdist} shows that the regions of enhanced magnetic pressure
associated with the isobaric modes
are numerous enough to dominate the total magnetic energy budget in the vicinity
of the midplane.  As much as half the magnetic energy can be located
in regions where the magnetic pressure is more than twice the local
horizontal average.  The maximum enhancement ratio in a single horizontal
slice can be as high as $\sim10$ at certain times and places.

\section{Radiative Damping}
\label{sec:silkdamping}

Any phenomenon involving radiation pressure fluctuations (acoustic or isobaric)
can be dissipated (i.e., lose its mechanical energy to heat, with associated
entropy production) via photon diffusion, which enters our formalism
in the diffusion equation.  In fact, we find that
both types of radiation pressure fluctuation contribute significantly to the
radiative damping, although the isobaric perturbations dominate.

\begin{figure*}
\plotone{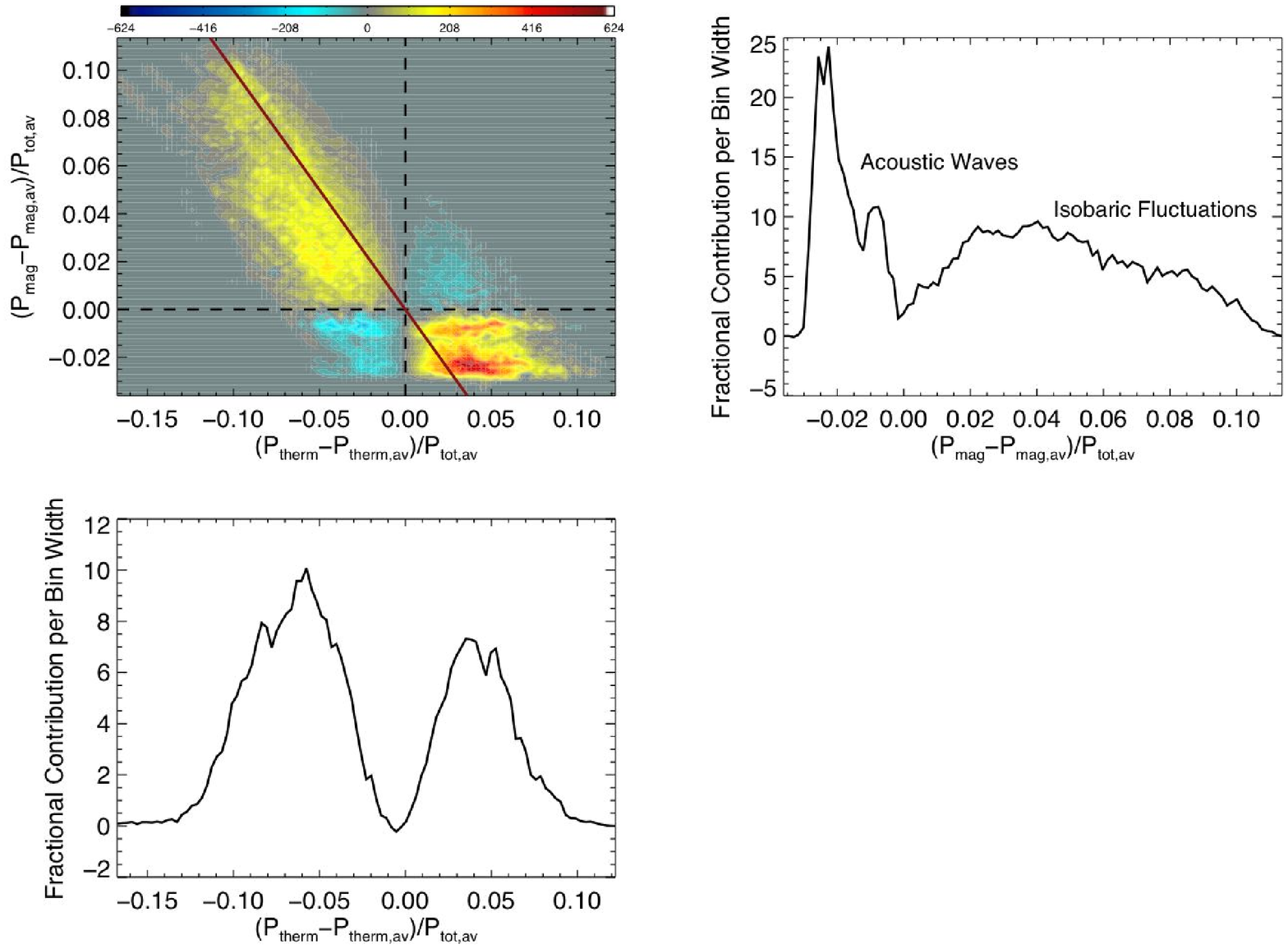}
\caption{Upper left:  two dimensional distribution function of the $z=0$
midplane radiative damping rate as a function of fractional thermal pressure
perturbation and fractional
magnetic pressure perturbation, averaged over the 200 to 202 orbit interval in
simulation 0519b.  The distribution is normalized such that the integral
over both perturbation variables is unity.  The red line shows the locus of
perfect isobaric perturbations.  Upper right:  projection of the distribution
of radiative damping across all thermal pressure perturbations, as a function of
magnetic pressure perturbation.  The regions largely associated with acoustic
waves and isobaric fluctuations are indicated.  Lower left:  projection of the
distribution of radiative damping across all magnetic pressure perturbations, as
a function of thermal pressure perturbation.
\label{fig:silkdist2d}}
\end{figure*}

One way of seeing this is to examine the integrated radiative damping as a
function
of fractional magnetic and thermal pressure fluctuations.  This is shown in the
upper left plot of Figure~\ref{fig:silkdist2d} for the 200 to 202 orbit
interval at the $z=0$ midplane in simulation 0519b.  (The corresponding
incidence of fractional magnetic and thermal pressure fluctuations themselves
was shown in the upper right hand plot of
Figure~\ref{fig:dpmagvsdpthermvsdrho}.)
Two distinct regions in this plane contribute to the radiative damping, one
associated with large positive magnetic fluctuations stretching up the isobaric
locus, and one associated with slightly negative magnetic fluctuations
that can occur with both positive and negative thermal pressure fluctuations
that are mostly due to acoustic waves.

Integrating the data of the 2-d distribution function over all thermal pressure
perturbation values results in the distribution shown in the upper right plot of
Figure~\ref{fig:silkdist2d}.   Because weak magnetic fields
(negative magnetic pressure perturbations) dominate the spatial volume in the
midplane, radiative damping from negative magnetic pressure perturbations here
is primarily due to acoustic waves.  In contrast, positive magnetic pressure
perturbations are mostly associated with the strongly magnetized, nonlinear
isobaric fluctuations.  The contributions to radiative damping from these two
types
of fluctuation are comparable:  35 percent for acoustic waves, and 65
percent for isobaric fluctuations for the 200-202 orbit epoch in simulation
0519b, for a total midplane radiative damping rate of
$9\times10^{14}$~ergs~cm$^{-3}$~s$^{-1}$.

Because the acoustic waves are much faster oscillating fluctuations, they are
presumably responsible for the rapid variability seen in the time dependence of
the box-integrated radiative damping (upper dashed gray curve) of
Figure~\ref{fig:dissstressvst0519b}.  In fact, the origin of the high
frequency spikes in the power spectrum of the box-integrated radiative damping
shown in Figure~\ref{fig:mpdvpowspec} is now clear.  All of these spikes
are higher order harmonics of a fundamental frequency of
$\simeq2.36$~inverse orbital periods.  This frequency is exactly the inverse
of the shear time of the box, and reflects the time interval between successive
pulses of the dominant $|n_y|=1$ nonaxisymmetric waves (see
Figure~\ref{fig:ptotwaveamplitudes}).

The measured contributions to the radiative damping can also be estimated
analytically from the observed fluctuation amplitudes, at least for the
nonaxisymmetric fast waves, which are not too nonlinear.  All the fluctuations
that we observe in the midplane regions have length scales such that the
photon diffusion time is much longer than the sound crossing time.
When radiation pressure dominates both magnetic and gas pressure,
the exponential damping rate of the amplitude of a linear, plane wave
fast mode (acoustic wave) in this slow diffusion limit is given by
\begin{equation}
\Gamma_{\rm acoustic}\simeq{k^2c\over6\bar{\kappa}^{\rm R}\rho},
\end{equation}
where $k$ is the wavenumber of the mode.
Physically, acoustic waves in a radiation dominated plasma are simply damped
at the rate that photons diffuse across a wavelength.
The acoustic radiative damping rate can therefore be estimated by multiplying
the average wave energy density by twice this damping rate (twice because
energy is proportional to the amplitude squared), i.e.
\begin{eqnarray}
Q_{\rm rad,acoustic}&=&\left<{(\delta P_{\rm tot})^2\over2c_{\rm t}^2\rho}+
{1\over2}\rho(\delta v)^2\right>(2\Gamma)\cr
&=&{(\delta P^2_{\rm tot,max})^2\over\rho c_{\rm t}^2}\Gamma\cr
&=&\left({k^2c \over 6\bar{\kappa}^{\rm R}}\right)\left({3c_{\rm t}\over4}
\right)^2 \left({\delta P_{\rm tot,max}\over
P_{\rm tot}}\right)^2,
\end{eqnarray}
where $c_{\rm t}\simeq6\times10^8$~cm~s$^{-1}$ is the total sound
speed in the radiation dominated plasma (see Figure~\ref{fig:speeds}).  From
Figure~\ref{fig:spiralwavesmidplaneplot}, we estimate a typical pressure
amplitude
of $\delta P_{\rm tot,max}/P_{\rm tot}\sim0.06$ and a wavenumber of
$k\sim2\pi/(0.3H)$.  For these numbers, we find
$Q_{\rm rad,acoustic}\sim3\times10^{14}$~ergs~cm$^{-3}$~s$^{-1}$, very close
to our measured acoustic contribution near this epoch of 35 percent of
$9\times10^{14}$~ergs~cm$^{-3}$~s$^{-1}$.

The short two-orbit epoch we have analyzed in detail in this section
is not atypical for this simulation: the midplane radiative damping rate of
$9\times10^{14}$~ergs~cm$^{-3}$~s$^{-1}$ at this time is close to the
midplane radiative damping rate time-averaged over the entire $\sim400$ orbits
of the simulation (upper gray dashed curve of Figure~\ref{fig:dissstress0519b}).
We reiterate here how significant these numbers are (as much as 22 percent of
the work done by the walls when integrated over the entire box) for the overall
energy budget at these high levels of radiation pressure support.

Although the contributions of acoustic waves and isobaric fluctuations to
radiative damping are comparable near the midplane, plots similar to
Figure~\ref{fig:silkdist2d} at fixed heights more than
$\simeq 0.5H$ away from the midplane clearly show that isobaric fluctuations
become dominant at these higher altitudes.  At these locations,
the two-dimensional distribution of radiative damping as a function
of magnetic and thermal pressure closely follows the isobaric locus for
both positive and negative magnetic perturbations, in
contrast to the midplane distribution (upper left plot of
Figure~\ref{fig:silkdist2d}).  This fact is consistent with the patterns of
fluctuations that we observe as a function of height in
Figure~\ref{fig:spiralwavesxzplot}.  In that figure, we see that the vertical
yellow and blue stripes near the midplane associated with nonaxisymmetric
acoustic waves in total pressure (upper left plot) extend out only to
$|z|\sim 0.2$--$0.3H$.  Beyond that, isobaric fluctuations of both signs of
thermal and magnetic pressure dominate the volume (lower left and right plots).

\section{The Nature of Vertical Advective Fluxes}
\label{sec:verticaladvection}
We now turn to examine two possible origins of the vertical radiation advection
energy flux: transport by trapped vertical waves and buoyancy.

\subsection{Vertical Epicyclic and Acoustic Waves}
\label{sec:verticalwaves}

\begin{figure}
\plotone{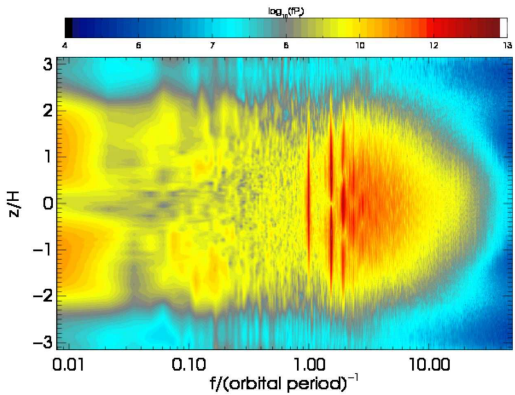}
\caption{Vertically resolved power spectrum of the horizontally-averaged
radiation advection flux $(Ev_z)_{\rm av}$ in simulation 0519b.}
\label{fig:eadpowspeczfreq}
\end{figure}

Figure~\ref{fig:eadpowspeczfreq} shows the temporal power spectrum of the
vertical radiation advection energy flux at every height $z$ in simulation
0519b.  A hierarchy of vertical modes is apparent at frequencies at and
above the orbital frequency.  All but the lowest frequency of these modes
are identical to those causing the oscillations and discrete frequency
peaks in the mechanical pumping shown in Figures~\ref{fig:dissstressvst0519b}
and \ref{fig:mpdvpowspec}.

The lowest order mode at the orbital frequency is just a vertical epicyclic
oscillation, excited by vertical asymmetries in momentum losses through the
vertical boundaries of the box. The vertical velocity $v_z$ in this mode is
the same at all heights, and the mode simply uniformly displaces the entire
plasma up and down in the box.  It produces zero compression
or rarefaction, and therefore does no pressure work, which explains its
complete absence in the temporal power spectra of pressure work terms shown
in Figure~\ref{fig:mpdvpowspec}.  Because it bodily displaces trapped photons
up and down, it modulates the radiation advection energy flux, but it produces
zero radiation advection {\it through} the plasma.

The next lowest frequency mode is a vertical breathing mode, and higher and
higher frequency modes clearly represent standing acoustic waves with
increasing numbers of vertical velocity nodes.  The vertical velocities of
these modes also modulate the radiation advection, but the net time-averaged
flux would be zero if these modes were purely adiabatic.
However, radiative damping of these modes does produce a net radiation
advection energy flux through the plasma over time.
With a small amount of analytic work (see appendix for details), we can
directly calculate this flux and compare it to the measured value.

The strongest mode is the breathing mode, and we measured its velocity
amplitude as follows.  First we
computed unwindowed, unbinned power spectra of the vertical velocity time
series at each height in the box.  We then summed the power at each height
over the measured mode line profile in these power spectra (corresponding
to the lowest peak in the upper green and black curves of
Figure~\ref{fig:mpdvpowspec}), and fit the resulting power at each height
in the midplane regions with the square of the expected mode eigenfunction
(linear for the breathing mode).  Finally, we used
Parseval's theorem to determine the actual velocity amplitude of the mode.
The results were $\simeq(1$~to~$7)\times10^6$~cm~s$^{-1}$ (roughly
one percent of the thermal sound speed) across the simulations at one pressure
scale height from the midplane.

The damping rate of the breathing mode is (see eq. \ref{eq:gammaradappendix})
\begin{equation}
\Gamma_{\rm rad}\sim{c\Omega^2\over\kappa_{\rm es} E_0},
\label{eq:gammarad}
\end{equation}
where $E_0$ is the radiation energy density at the midplane.  Not
surprisingly, this is approximately the reciprocal of the nominal
thermal time of the disk.\footnote{The actual thermal time as measured by
the time average of the box-averaged thermal energy content divided by the
sum of the top and bottom emergent radiation fluxes is several times shorter
than this nominal thermal time, presumably because the dissipation profiles
peak off the midplane and some of the energy is carried outward by mechanical
motions (see discussion around eq. A8 of \citealt{hirbk09}).}  Because the
breathing mode is the longest
vertical wavelength acoustic mode, it should be damped at a rate given
roughly by the rate at which photon diffusion can cool the disk.

The radiative damping of the breathing mode causes phase changes between the
velocity and pressure perturbations, which result in a secular flux of
mechanical energy carried by the mode.  This energy flux is given
by (see eqs. \ref{eq:dvdpthappendix} and \ref{eq:dvdpmagappendix} in the
appendix)
\begin{equation}
\left<\delta v\left[\Delta P_{\rm therm}+
\Delta\left({B^2\over8\pi}\right)\right]
\right>={3z\Gamma_{\rm rad}A_{\rm b}^2\over14\Omega^2H_P^2}
\left({4E\over9}+{B^2\over4\pi}\right),
\label{eq:waveflux}
\end{equation}
where $A_{\rm b}$ is the velocity amplitude of the breathing mode at one
pressure scale height $H_P$ from the midplane.

Evaluating equations (\ref{eq:gammarad}) and (\ref{eq:waveflux}), we find
a peak radiation advection flux of only
$7\times10^{16}$~ergs~cm$^{-2}$~s$^{-1}$ at the breathing mode
amplitude observed in the 0519b simulation; the other simulations
have comparable fluxes, although the ones with lower radiation to gas
pressure ratios are smaller by factors of several.  However,
Figure~\ref{fig:energyflux}
shows that the actual time-averaged peak radiation advection flux in 0519b
is more than five orders of magnitude
larger:  $1-2\times10^{22}$~ergs~cm$^{-2}$~s$^{-1}$.  The breathing mode
amplitude would have to be at least $\sim 400$ times larger,
i.e. several times larger than the sound speed,
in order to produce that large a flux.  Similar discrepancies exist in
all six simulations, and we therefore conclude that the acoustic waves
do not contribute significantly to the time-averaged outward radiation
advection.  Instead, they simply modulate it at high frequencies.

Similar calculations can be used to estimate the dissipation rate due to
radiative damping of the breathing mode.  For example, in simulation 0519b,
this is $\simeq4\times10^{10}$~ergs~cm$^{-3}$~s$^{-1}$ at the midplane, orders
of magnitude smaller than the actual rates of radiative damping and pressure
work that we find in the midplane regions of this simulation
(Figure~\ref{fig:dissstress0519b}).  The same is true for all the other
simulations, and we conclude that the standing vertical acoustic waves
contribute negligibly to radiative damping and the time-averaged mechanical
pumping work, although they do significantly modulate the latter.

\subsection{Buoyancy as the driver of advection}
\label{sec:buoyancy}

\begin{figure}
\plotone{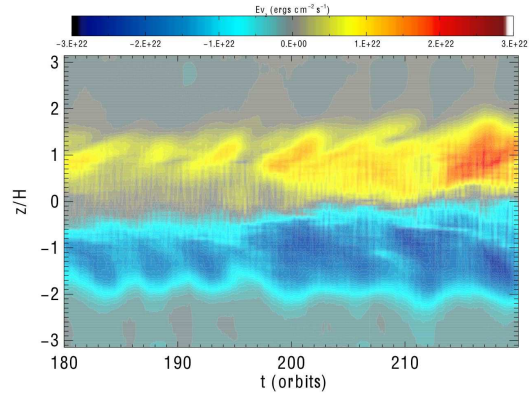}
\caption{Horizontally-averaged radiation advection flux $(Ev_z)_{\rm av}$,
smoothed over a running two orbit interval, as a function of height $z$ and
time $t$ in simulation 0519b. Fine scale vertical striations in this plot
are what remain of the vertical epicyclic and acoustic modulation of the
radiation advection after the two-orbit smoothing.}
\label{fig:eadsmoothspacetime}
\end{figure}

\begin{figure}
\plotone{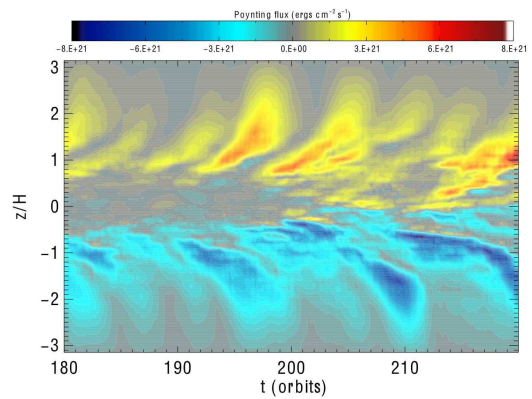}
\caption{Horizontally-averaged Poynting flux,
smoothed over a running two orbit interval, as a function of height $z$ and
time $t$ in simulation 0519b.}
\label{fig:poynsmoothspacetime}
\end{figure}

We have just seen that, while the radiation advection is modulated by
vertical acoustic modes, these waves do not contribute
significantly to the net radiation advection flux.  If we smooth over this
rapid modulation, a clear pattern emerges as shown in
Figure~\ref{fig:eadsmoothspacetime}.  The radiation advection flux is modulated
on long ($\sim5$ orbit) time scales, and this modulation is in fact highly
correlated with a similar modulation in the smoothed spacetime pattern
of vertical Poynting flux shown in Figure~\ref{fig:poynsmoothspacetime}.  These
spacetime patterns of Poynting flux have been seen in all previous vertically
stratified MRI simulations, both with and without explicit thermodynamics
\citep{bra95,sto96,mil00,hir06,dav10,shi10}, but their origin has remained a
mystery.

We argue here that both the radiation advection and the vertical Poynting
flux are due to buoyancy in the region inside $|z| \sim H$.
That there is any buoyancy at all in the midplane regions is at first
surprising, as the time and horizontally averaged pressure and density
profiles are linearly stable to buoyant instabilities in the midplane regions.
Indeed, when the square of the hydrodynamic Brunt-V\"ais\"al\"a frequency,
\begin{equation}
N^2=g\left({1\over\Gamma_1}{d\ln P_{\rm therm}\over dz}-
{d\ln\rho\over dz}\right),
\end{equation}
is computed using horizontally- and time-averaged pressure and density
profiles, it
is positive everywhere off the midplane in all our simulations, indicating a
hydrodynamically stable average vertical entropy profile.

Magnetic fields can still cause buoyancy instabilities in
the form of the interchange and undulatory Parker modes.  The interchange
mode is stable provided the ratio of magnetic field strength to mass density
does not decrease outward too fast (e.g. \citealt{ach79}).  In our
simulations, the time-averaged profiles have $d/dz\ln(B/\rho)>0$ everywhere
off the midplane, so the interchange mode is linearly stable.

\begin{figure}
\plotone{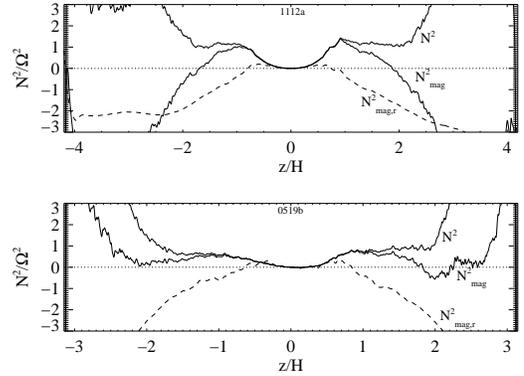}
\caption{Square of magnetic Brunt-V\"ais\"al\"a frequency appropriate for
the undulatory Parker instability, as computed from the time averaged pressure
and density profiles of simulations 1112a (top) and 0519b (bottom). The
solid curves are for slow radiative diffusion (first equality of
eq.~\ref{eq:bruntnewcomb}), while the dashed curves are for rapid
radiative diffusion (eq.~\ref{eq:bruntgilman}).  (The second equality of
equation \ref{eq:bruntnewcomb} produces curves that deviate from the solid
curves only in the surface regions where magnetic tension forces are
significant.)}
\label{fig:brunts}
\end{figure}

This leaves the undulatory Parker modes, which are typically linearly
unstable in the surface layers \citep{bla07}.  In the midplane regions,
where radiative diffusion is slow, the linear stability criterion
\citep{new61} can be expressed in terms of the requirement that
the square of the magnetic Brunt-V\"ais\"al\"a frequency be positive.  This
can be defined as
\begin{equation}
N_{\rm mag}^2\equiv g\left(-{g\over c_{\rm t}^2}-{d\ln\rho\over dz}\right)
             =N^2+{gv_{\rm A}^2\over c_{\rm t}^2}{d\ln B\over dz},
\label{eq:bruntnewcomb}
\end{equation}
where $c_{\rm t}\equiv(\Gamma_1P_{\rm therm}/\rho)^{1/2}$ is the adiabatic sound
speed in the gas plus radiation mixture, $v_{\rm A}$ is the Alfv\'en speed, and
the last equality follows if the field is purely horizontal with no vertical
tension forces.  The solid lines in Figure~\ref{fig:brunts} show that the
midplane regions are stable to the undulatory Parker modes by this criterion.
If radiative diffusion is fast enough to suppress temperature
fluctuations, the undulatory Parker stability criterion \citep{gil70}
can instead be written in terms of a modified magnetic Brunt-V\"ais\"al\"a
frequency
\begin{equation}
N_{\rm mag,r}^2\equiv{gv_{\rm A}^2\over(c_{\rm i}^2+v_{\rm A}^2)}
{d\ln B\over dz},
\label{eq:bruntgilman}
\end{equation}
where $c_{\rm i}\equiv(p/\rho)^{1/2}$ is the isothermal sound speed in the gas.
The dashed lines in Figure~\ref{fig:brunts} show the profiles of
$N_{\rm mag,r}^2$.  Hence even
if radiative diffusion were fast in the midplane regions, they would
still be stable simply because the time and horizontally-averaged magnetic
pressure typically {\it increases} outward in these regions.

The midplane regions, {\it viewed in terms of horizontally-averaged quantities},
are therefore linearly stable to buoyant perturbations.  Nevertheless,
finite amplitude perturbations with three dimensional structure within the
turbulence can still be locally buoyant.
As we have already seen, magnetosonic slow modes create regions of high
magnetic pressure and low gas density.  Here they have nonlinear amplitude,
but retain these qualitative characteristics.  Because they have the
same pressure as other locations at that altitude, but have lower density,
they are, of course, buoyant.  Such low density, high magnetic pressure
fluctuations are clearly present throughout the region within $\pm 0.7H$ of
the midplane shown in Figure~\ref{fig:spiralwavesxzplot}.

\begin{figure*}
\plotone{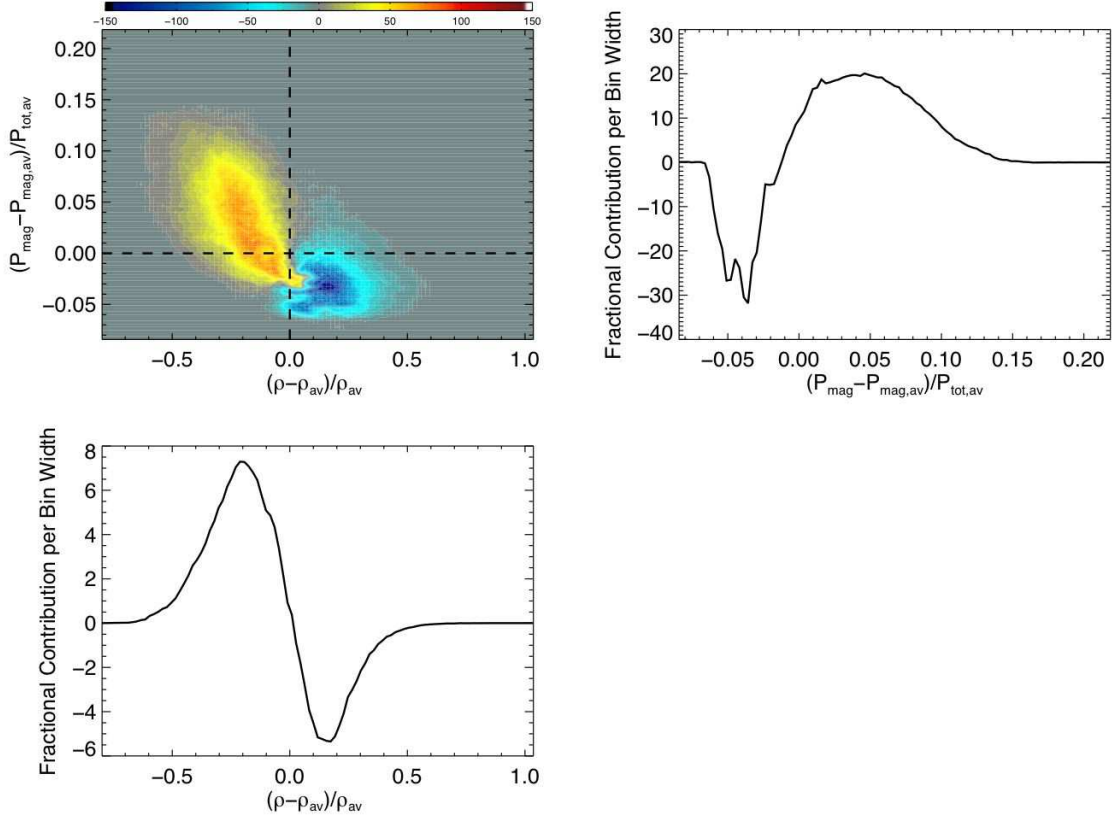}
\caption{Upper left: two dimensional distribution of the 200 to 208 orbits
time averaged radiation advection at $z=+0.5H$ in simulation 0519b, as
a function of scaled magnetic pressure perturbation and density perturbation.
The distribution is normalized such that the two dimensional integral over
both perturbation variables is unity.  Upper right: projection of the radiation
advection distribution across all density perturbations, as a function of
magnetic pressure perturbation.  Lower left:  projection of the radiation
advection distribution across all magnetic pressure perturbations, as a
function of density perturbation.
\label{fig:radadvect}}
\end{figure*}

Unlike the breathing modes, these fluctuations {\it do} carry enough energy flux
to explain the advection.  Figure~\ref{fig:radadvect} shows the two
dimensional distribution of radiation advection flux at $z=+0.5H$ in
simulation 0519b, averaged over 200 to 208 orbits, as a function of
magnetic pressure and density perturbations.  Integrating over all densities
and magnetic fields to compute the total net radiation advection gives
$9\times10^{21}$~ergs~cm$^{-2}$~s$^{-1}$, which is of course consistent with
the horizontally and time-averaged radiation advection at this height around
this epoch (Figure~\ref{fig:eadsmoothspacetime}). It is also consistent
with the average radiation advection flux that we see at this height over
the entire duration of the simulation (Figure~\ref{fig:energyflux}).

Most importantly, it is clear that magnetic pressure and density continue to
be anti-correlated above the midplane, just as they are in the midplane.
Moreover, outward radiation
advection tends to be associated with low density and high magnetic pressure,
and inward radiation advection tends to be associated with high density and
low magnetic pressure.  This can also be seen if we integrate over all
densities (the distribution in the upper right plot) or all magnetic
pressures (the distribution in the lower left plot).  Thus, the advection
is clearly associated with these local regions of buoyancy.

A corollary of this finding is that the regions responsible for the bulk
of upward radiation advection actually have radiation energy density somewhat
{\it smaller} than other regions at the same altitude.  In other words,
this process, although it bears some resemblance to classical convection,
differs strikingly in that it carries heat upward in fluid elements that
are initially {\it cooler} than their surroundings.  Instead of being due to a
higher entropy per unit mass (which in fact they have in spite of being cooler
than their surroundings), their buoyancy is driven by a higher
magnetic pressure per unit mass.

The fact that cooler than average buoyant fluid elements can still give rise
to a net outward energy flux can also be seen by the following simple
argument.\footnote{We thank the referee for suggesting this argument to us.}
In any given horizontal plane, divide
the radiation energy density, mass density and vertical velocity into their
horizontal average plus
perturbations (just as we did with other variables to separate radiative
damping from mechanical pumping in the pressure work terms in section
\ref{sec:thermodynamics} above):  $E=E_{\rm av}+\delta E$,
$\rho=\rho_{\rm av}+\delta\rho$, and $v_z=v_{z,{\rm av}}+\delta v_z$.  We
observe zero net vertical mass flux in the simulations, so
$v_{z,{\rm av}}=-(\delta\rho\delta v_z)_{\rm av}/\rho_{\rm av}$.  The
horizontal average of the radiation advection energy flux is therefore
given by
\begin{equation}
{(Ev_z)_{\rm av}\over E_{\rm av}}=\left[\delta v_z
\left({\delta E\over E_{\rm av}}-{\delta\rho\over\rho_{\rm av}}\right)
\right]_{\rm av}
\end{equation}
The fractional density fluctuations are
generally much larger in magnitude than the fractional temperature (or
radiation pressure or energy density) fluctuations, as we can see in
Figures~\ref{fig:spiralwavesmidplaneplot} and \ref{fig:spiralwavesxzplot}.  It
is the sign of these density fluctuations that give rise to an overall
outward mean radiation advection energy flux.  Put another way, the underdense
regions give rise to a net outward horizontal average velocity, and because
the temperature fluctuations are so small, this net outward velocity times
the average radiation energy density is approximately the net outward
radiation advection flux.

While the rising magnetized fluid parcels start out cooler than their
surroundings, they nevertheless eventually become hotter than their
surroundings, and it is
for this reason that they are able to transfer heat from the hot midplane
regions to the cooler regions away from the midplane.  We can see this
qualitatively through the following linearized treatment.
Consider a bundle of (mostly horizontal) field lines that rises
an infinitesimal height $\Delta z$ from some initial height $z$.  We take the
gas pressure to be negligible, and assume that radiative diffusion is slow so
that the plasma inside the bundle undergoes an approximately adiabatic change
(as shown by Fig.~\ref{fig:speeds}, in this region the diffusion speed on a
lengthscale $0.1H$ is about comparable to the typical upward advective speed
in this region).  We further assume that the mean gradient in the total pressure
(radiation plus magnetic) is in hydrostatic balance with gravity.  Pressure
equilibrium with the surroundings then implies that the radiation pressure
$P_{\rm rad}^\star$ of the plasma {\it inside} the bundle will change by
\begin{equation}
\Delta P_{\rm rad}^\star={-\rho g\Delta z\over1+3P_{\rm mag}^\star/
2P_{\rm rad}^\star},
\end{equation}
where $P_{\rm mag}^\star$ is the initial magnetic pressure inside the bundle.
On the other hand, the radiation pressure in the local background will change
by
\begin{equation}
\Delta P_{\rm rad}={-\rho g\Delta z\over1+dP_{\rm mag}/dP_{\rm rad}},
\end{equation}
where $dP_{\rm mag}/dP_{\rm rad}$ is a derivative following background pressures
as they vary with height.  We therefore find that the rising fluid parcel will
cool {\it less} rapidly with height than the background provided
\begin{equation}
\left({P_{\rm mag}\over P_{\rm mag}^\star}\right)\left({P_{\rm rad}^\star\over
P_{\rm rad}}\right){d\ln P_{\rm mag}\over d\ln P_{\rm rad}}<{3\over2}.
\end{equation}
This is nearly always true in our simulations, as the bundle starts out
more magnetized and cooler than its surroundings, and the background magnetic
pressure generally falls less rapidly with height than the radiation
pressure (see e.g. Fig. 16 of \citealt{hir09}).
In fact, near the midplane, the magnetic
pressure generally rises with height, while the radiation pressure
falls.  For example, in the advective region of 0519b ($|z| \lesssim 1.5 H$),
$d\ln P_{\rm mag}\over d\ln P_{\rm rad}$ rises outward from $\simeq -1.5$ very
near the midplane to a plateau at $\simeq 0.5$.  Thus, rising magnetized
fluid parcels eventually become hotter than their surroundings, at which point
heat will be transferred outward to the surroundings by radiative diffusion.

\begin{figure*}
\plotone{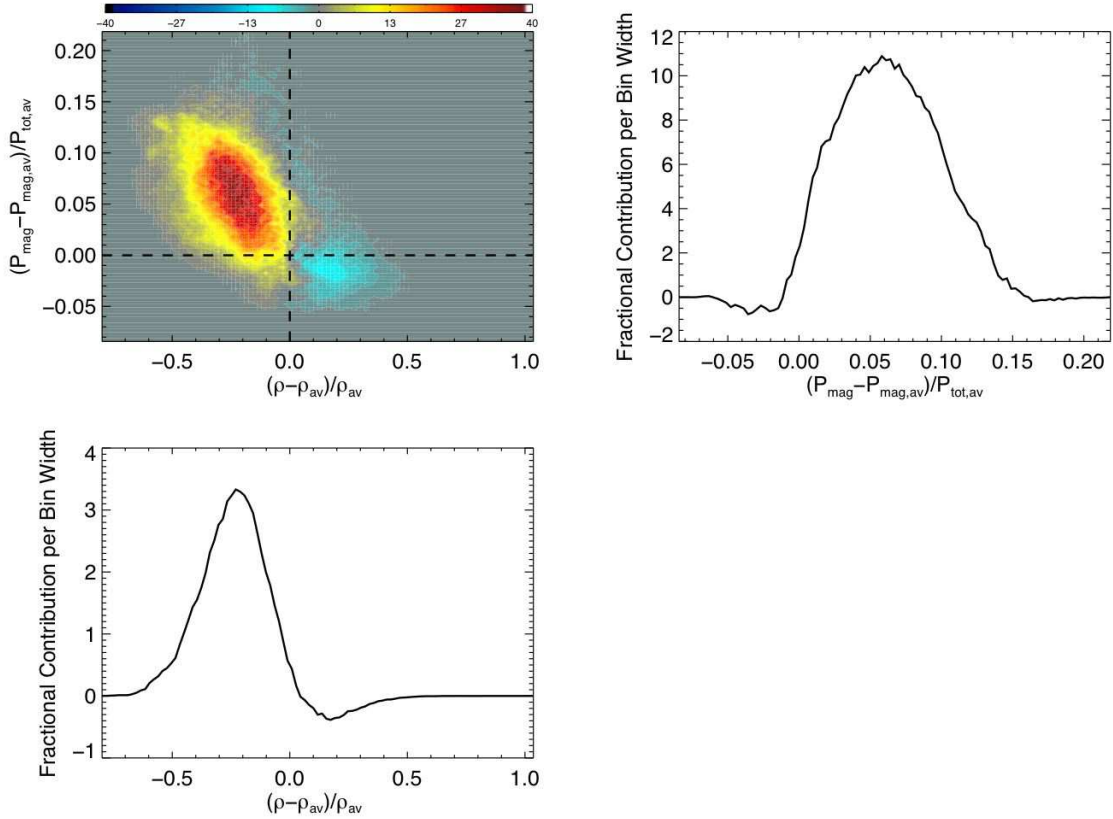}
\caption{Upper left: two dimensional distribution of the 200 to 208 orbits
time averaged Poynting flux at $z=+0.5H$ in simulation 0519b, as
a function of scaled magnetic pressure perturbation and density perturbation.
The distribution is normalized such that the two dimensional integral over
both perturbation variables is unity.  Upper right: projection of the Poynting
flux distribution across all density perturbations, as a function of
magnetic pressure perturbation.  Lower left:  projection of the Poynting
flux distribution across all magnetic pressure perturbations, as a
function of density perturbation.
\label{fig:poyndist}}
\end{figure*}

Not surprisingly, regions of strong Poynting flux are even more strongly
associated with high magnetic field, low mass density isobaric fluctuations,
as shown in Figure~\ref{fig:poyndist}.  This figure is constructed in a
fashion exactly analogous to Figure~\ref{fig:radadvect}
illustrating radiation advection.
Supporting our argument that the two are fundamentally the same, the patterns
seen in this figure are very similar to those seen in the radiation advection
figure.  The principal contrast is that, unlike the radiation advection case,
there is very little downward Poynting flux; it is almost exclusively upward.
This follows, of course, from the fact that the upward-moving isobaric
fluctuations have {\it greater} than average magnetic energy density,
whereas they have smaller than average radiation energy density.

\subsection{Characteristic speed of advective motions}

\begin{figure}
\plotone{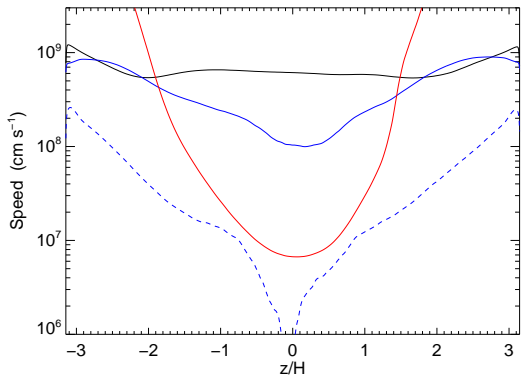}
\caption{Time- and horizontally-averaged vertical profiles of thermal sound
speed (black), Alfv\'en speed (solid blue), radiation diffusion speed over
a length scale of $0.1H$ (solid red), and magnetically-weighted upward
advective speed (dashed blue), for simulation 0519b.
\label{fig:speeds}}
\end{figure}

Because the outward advection fluxes are associated with regions of
enhanced magnetic field, we can define a horizontally-averaged vertical
advection speed by using the $z$-component of the Poynting vector
${\bf S}=c({\bf E}\times{\bf B})/(4\pi)$,
\begin{equation}
v_{\rm adv} \equiv 4\pi{\int_{-L_x/2}^{L_x/2} dx \int_{-L_y/2}^{L_y/2} dy
S_z(x,y,z)\over\int_{-L_x/2}^{L_x/2} dx \int_{-L_y/2}^{L_y/2} dy
B^2(x,y,z)}.
\end{equation}
We compare the time-averaged vertical profile of this speed in simulation 0519b
to the time-averaged vertical profiles of the thermal sound speed $c_{\rm t}$
and the Alfv\'en speed in
Figure~\ref{fig:speeds}.\footnote{This figure shows two other interesting
features that are outside the scope of the current discussion.  First,
the thermal sound speed is remarkably constant with height, a property that
is shared by all the simulations we analyze in this paper.  Second, the
Alfv\'en speed is not very much larger than the thermal sound speed in the
surface layers in simulation 0519b.  This is in marked contrast to all the
other simulations that have lower radiation to gas pressure ratio, which have
very strongly magnetized surface regions that are Parker unstable
\citep{bla07}.  Parker instability dynamics is still apparent in the surface
regions of 0519b, but it is conceivable that at still higher levels of
radiation support that magnetic support and Parker dynamics would become less
important.}
The advection speed is several percent of the sound speed
in the near-midplane
region where advection is most important, but rises outward, reaching Mach
numbers a little less than unity in the outer layers of the disk.
This outward increase in $v_{\rm adv}$
is consistent with the characteristic buoyant rise time of 5--10 orbits
for magnetic features: because $H \sim c_{\rm t}/\Omega$, such a
characteristic time implies a mean upward Mach number $\sim 0.1$--0.2.

Figure~\ref{fig:speeds} also shows the characteristic one-dimensional
speed of radiation diffusion $v_{\rm diff}$ over a length scale of
$\ell=0.1H$, the characteristic size of the buoyant fluctuations
(Figure~\ref{fig:spiralwavesxzplot}).  We define this speed as this length
scale divided by the radiation diffusion time, i.e.
\begin{equation}
v_{\rm diff}\equiv{c\over3\ell\bar{\kappa}^{\rm R}\rho}.
\end{equation}
The diffusion speed increases rapidly away from the midplane as the density
drops.  It is comparable to the advection speed in the midplane regions,
and then greatly exceeds the advection speed at heights greater than
$\simeq H$.  This
coincides with the heights above which radiative diffusion finally becomes
dominant over radiation advection in this simulation (bottom panel of
Figure~\ref{fig:energyflux}).

The average advection velocity profile also allows us to estimate the work done
by the plasma in driving radiation advection.  Figure~\ref{fig:speeds} shows
that at $z=H$, $v_{\rm adv}\simeq10^7$~cm~s$^{-1}$ in simulation 0519b.
The average radiation pressure at that height is
$\simeq8\times10^{14}$~dyne~cm$^{-2}$.  Hence the mechanical pumping work
associated with radiation advection at this height is
$\simeq-(E/3)dv_{\rm adv}/dz\sim-(E/3)v_{\rm adv}/H\sim
-2\times10^{15}$~erg~cm$^{-3}$~s$^{-1}$.  This agrees very well
with the actual mechanical pumping in Figure~\ref{fig:dissstress0519b} (lower
gray dashed curve).  We conclude that the vertical expansion motions
associated with radiation advection are largely responsible for the mechanical
pumping work done by the plasma.

\section{Discussion}
\label{sec:discussion}

We have now answered a number of the questions posed at the beginning of this
paper on the thermodynamics of radiation pressure dominated disks.  Central
to our discussion is the existence of high magnetic energy density, low gas
density regions that manifest within the turbulence as nonlinear slow
magnetosonic modes.  We named these regions ``isobaric fluctuations" because
the total pressure perturbation within them (thermal plus magnetic) is
quite small.  Higher than average magnetic pressure is associated with lower
than average thermal pressure.  The isobaric fluctuations are nonlinear, in
that they are associated with order-unity magnetic intermittency.  The
magnetic energy density in these fluctuations is a few times the volume
mean, and they constitute a significant fraction of the magnetic energy in the
turbulence.  On the other hand, these regions are rarely of
sufficient size and coherence that they can be thought of as ``flux tubes".

We have elucidated the nature and role of radiative damping.
To supply the energy that is ultimately dissipated by
photon diffusion, thermal pressure does work {\it on} the fluid in association
with compressive fluctuations.  This can be distinguished from work done
{\it by} the fluid in driving vertical mechanical motions by subtracting off
appropriate horizontal averages in presure and velocity
(equation~\ref{eq:pressureworksplit}).  The compressive fluctuations
responsible for radiative damping are a combination of
isobaric fluctuations and acoustic waves.  In the midplane regions of
our most radiation dominated simulation, we find that the dissipation
associated with radiative damping is
roughly 65\% due to isobaric modes and 35\% to acoustic waves.  At higher
altitude, isobaric fluctuations dominate the dissipation associated with
radiative damping.

It is noteworthy that the time-averaged spatial distribution of radiative
damping is similar in shape to the time-averaged spatial distribution of
magnetic and kinetic energy dissipation (Figures~\ref{fig:dissstress1112a} and
\ref{fig:dissstress0519b}).  Similarly, the box-averaged radiative damping is
well correlated with the magnetic and kinetic energy
dissipation rates as a function of time: averaging over the short time scale
fluctuations associated with acoustic waves in the radiative damping (upper
dashed gray) curve of Figure~\ref{fig:dissstressvst0519b} makes this history's
relation to the magnetic (blue) and kinetic energy (green) dissipation curves
clear.  In addition, as we have already noted, a significant fraction of the
total magnetic energy is found in regions where the magnetic energy density is
of order twice the mean, places where strong isobaric fluctuations have led
to field concentration.  We therefore suspect that isobaric
fluctuations play a significant role in the grid scale magnetic and kinetic
energy dissipation.  In the magnetic case, this is perhaps not surprising, as
the large concentrations of magnetic field must be associated with sharp
gradients of magnetic field, as well as high electric current densities.

Surprisingly, we have also seen that when the ratio of radiation pressure
to gas pressure is large enough, the total dissipation rate due to radiative
damping can become comparable to the gridscale loss of magnetic energy.
If this trend continues to still larger radiation pressure, it will mean
that we will have the unique advantage of being able to resolve the principal
scale of dissipation in astrophysical MHD turbulence.  That will finesse
a great many of the uncertainties and difficulties that stand in the way
of a clear interpretation of MHD turbulence in many other contexts.

Indeed, there is growing evidence that the saturated state of MRI turbulence
may depend on the microscopic viscosity and resistivity of the plasma
\citep{fplh07,les07,sim09,dav10}.  In particular, much attention has
been paid to the sustainability and properties of the turbulence for
different magnetic Prandtl numbers (the dimensionless ratio of
viscosity to resistivity), because this sets the relative
sizes of the velocity and magnetic dissipation scales.
The inner, radiation dominated regions of optically thick, geometrically
thin accretion disks around black holes generally have magnetic Prandtl
number greater than unity if the viscosity and resistivity
are dominated by Coulomb scattering \citep{bal08}.  In other words, the
viscous scale, where velocity fluctuations are damped, exceeds the resistive
scale where magnetic fluctuations are damped.  Insofar as radiative damping
also suppresses velocity fluctuations, albeit only those which are
compressive, it reinforces this ordering: velocity fluctuations are damped
on larger spatial scales than magnetic fluctuations in radiation dominated
accretion disks.

Perhaps the most important question that we posed at the beginning of this
paper was how a radiation dominated disk handles a dissipation
rate that exceeds the critical value of $Q^\star=c\Omega^2/\kappa$ associated
with hydrostatic and radiative equilibrium.  As shown in
Figures~\ref{fig:dissstress1112a} and \ref{fig:dissstress0519b},
such an excess always occurs in the midplane regions.
We find that the radiative diffusion flux
is indeed constrained by hydrostatic equlibrium to have a divergence equal
to $Q^\star$, and excess energy flux is carried outward by radiative advection,
fluid motion carrying trapped photons.  A smaller (by $\sim\beta^{-1}$) energy
flux is conveyed electromagnetically by the same fluid motions.
At higher altitudes, the vertical energy flux
becomes almost entirely diffusive radiation flux, simply because it
becomes progressively harder for small fluid elements to trap photons as the
density falls.

The existence of substantial vertical energy flux carried by
something other than radiative diffusion or Poynting flux is a significant
addition to the repertory of accretion disk physics.  This is the first
time that advected radiation flux has been shown to be important in disks.

We emphasize that these advected energy fluxes are {\it not} classical
convection, as the horizontal and time-averaged entropy
profiles are stably stratified.  This is also true of the horizontal
and time-averaged magnetic profiles, at least in the midplane regions.
These advected fluxes are also not in any way related
to the old argument that radiation dominated disks have constant density
and are therefore convectively unstable.  As we discussed in the introduction,
this argument rested on the assumption that the dissipation per unit {\it mass}
is constant.  This is simply not true, nor is the density constant
with height.

{\it Local} buoyancy is the mechanism responsible for driving both
radiation advection and upward Poynting flux, and this buoyancy arises
from the low densities of the nonlinear, highly magnetized isobaric
fluctuations.  Because the lower than average thermal pressure within these
fluctuations is mostly radiation pressure, these buoyant
fluid parcels are cooler than their surroundings, at least initially.
Nevertheless, they
advect radiation internal energy outward because their magnetic pressure
makes them buoyant.  The work done by the plasma in driving this local
vertical expansion explains the mechanical pumping work that we separated
from the radiative damping in the total $PdV$ work.

Although these buoyant regions play an important role in the energy budget
in very radiation-dominated disks, they are also important for a different
reason in disks with weaker radiation content.  The same isobaric fluctuations
can also explain the characteristic upwelling of magnetic flux seen in
{\it all} stratified
disk simulations.  Radiation forces {\it per se} are unnecessary:
the only prerequisite is that there be thermal pressure fluctuations
accompanied by proportionate density fluctuations that are
balanced by opposite sign fluctuations in the magnetic pressure.  Data from
our earlier gas-dominated simulations show essentially the same dynamics as
the data from radiation-dominated simulations that has played the primary role
in this paper's analysis.

The quasi-periodic build up of magnetic flux in the midplane regions is also
associated with azimuthal field reversals, clearly indicating some sort
of dynamo activity \citep{bra95,joh09,dav10,gre10,one10,shi10}.
The fact that the resulting upwelling motions are energetically significant
is the first example that we know of in astrophysics of a magnetic dynamo
that is important for the global energetics of the medium.  For example,
the solar dynamo is in part driven by the outward energy transport associated
with convection in the sun.  In the case of a radiation-dominated accretion
disk, the dynamo creates buoyant high magnetic field regions which {\it then}
play a significant role in energy transport by advecting radiation
(and magnetic field) outward.  We have seen that this radiation advection is
just as important as radiative diffusion in the midplane regions for
the most radiation dominated simulation we have conducted so far.

A final outstanding question is how the turbulence ``knows" that it needs
to produce a heating rate that is of order the critical value $Q^\star$
required by hydrostatic and radiative equilibrium.  We have seen that, in
fact, when the overall radiation to gas pressure ratio is high,
the dissipation rate generally exceeds $Q^\star$ in the midplane
regions, and the excess heat is carried outward by radiation advection,
not radiative diffusion.  But is this radiation advection therefore somehow
regulated to match the requirements of overall thermal equilibrium?  We
suspect that the answer is no, at least not directly.

The nonlinear isobaric fluctuations are an inherent feature of intermittency
in MRI turbulence and are related (somehow) to the dynamo process producing
quasiperiodic reversals in the azimuthal field.  They are inherently
buoyant, and therefore the resulting rate of radiation advection is simply
proportional to how much radiation energy density is present in the fluid
to be carried upward with the magnetic field.  As we just discussed, these
fluctuations do appear to be associated with grid scale dissipation that
helps determine the local radiation energy density, but that is a {\it
slow} process.  Dissipation, which is still mostly magnetic in nature, has
to occur over time scales comparable to the thermal time to build up enough
heat to significantly change the radiation energy density, simply because
the volume-averaged magnetic energy density is so small.\footnote{This
inherent time lag is ultimately why the disk is thermally stable in the
radiation dominated regime \citep{hir09}.}

Instead, any small excess radiation energy density that appears because
the average dissipation exceeds $Q^\star$ results in an excess vertical
diffusive radiation flux.  The resulting excess radiation pressure force
quickly (on the fast sound crossing time scale) produces a vertical
expansion of the medium.  As a consequence, work is done by the fluid and
this lowers the radiation energy density, restoring hydrostatic equilibrium.
This mechanism may in part be responsible for the excitation of the standing
vertical acoustic waves.
It may very well be that hydrostatic equilibrium cannot be
so maintained if the turbulent dissipation becomes significantly stronger.
Outflows may appear at levels of radiation dominance we have not yet
explored in our simulations.

Ultimately, the mechanism that sets the midplane dissipation to be in excess
of $Q^\star$, and the level of ambient radiation energy density that is
then advected outward by buoyant magnetic field, is the level of turbulence
in the medium.  However, what sets the saturation level of MRI turbulence is
still an open question.

Finally, we remark that photon bubbles have been proposed as an additional
piece of physics that may alter the vertical transport of energy at high
luminosities \citep{beg06}.  The characteristic length scale of these
instabilities is $\sim c_{\rm i}^2/g$, where $c_{\rm i}$ is the isothermal
{\it gas} sound speed and $g=\Omega^2|z|$ is the acceleration due to gravity
\citep{gam98,beg01,bla03,tur05}.  In simulation 0519b, this is less than the
size of our grid zones at distances more than $\simeq0.5H$ away from the
midplane.  Clearly these
instabilities cannot be resolved in our simulations, and their possible
impact on the thermodynamics remains a topic for future study.

\section{Conclusions}
\label{sec:conclusions}

The thermodynamics of a radiation dominated disk differs significantly
from that envisaged in standard static models of accretion disks.  In addition
to microscopic dissipation associated with the MRI turbulent
cascade, radiative damping of compressive fluctuations plays an increasingly
significant role as the radiation to gas pressure ratio increases.  The
compressive fluctuations associated with this damping consist of fast
magnetosonic waves as well as nonlinear isobaric fluctuations arising from
the turbulence itself.  This damping is energetically significant: on the
order of tens of percent of the overall dissipation at the highest levels
of radiation pressure support that we have simulated.  It is also numerically
resolvable, in contrast to microscopic resistive and viscous dissipation.

Buoyancy of the highly magnetized, low density isobaric fluctuations is
responsible in part for the butterfly diagram that is always seen in
simulations of MRI turbulence with vertical gravity.  This buoyancy is
intrinsically three-dimensional in nature; even though the horizontally-averaged
structure is very stably stratified in the midplane regions, localized
concentrations of magnetic field generated by the turbulent dynamo are
underdense and still produce outward
advection of magnetic field.   When the plasma is radiation-dominated, these
buoyant motions become significant for the overall energetics of the plasma.
Outward advection of photons becomes comparable to radiative diffusion, and
the associated vertical expansion work must be included in balancing vertical
heat transport and dissipation.

This work was supported NSF grants AST-0707624 and AST-0507455, and
by a Grant-in-Aid for Scientific Research (No. 20340040) from the
Ministry of Education, Culture, Sports, Science and Technology of Japan.
Numerical computations were carried out partially on the Cray XT4 at
the Center for Computational Astrophysics at the National
Astronomical Observatory of Japan, and on the SX8 at the Yukawa Institute
for Theoretical Physics at Kyoto University.
Our work has benefited from conversations with numerous people, including
Steve Balbus, Shane Davis, John Hawley,
Tobias Heinemann, Ken Henisey, Gordon Ogilvie, Chris Reynolds,
Prateek Sharma, Jim Stone, Ted Tao, Neal Turner, and Ethan Vishniac.
We also thank the referee, Jeremy Goodman, for detailed constructive
criticism that greatly improved the paper.
Part of this work
was completed while OB was visiting the Kavli Institute for Astronomy and
Astrophysics in Beijing.

\appendix
\section{Vertical Energy Transport by Standing Vertical Modes}

In this appendix we outline the derivations of the damping times and vertical
energy fluxes of the standing vertical acoustic waves discussed in 
section~\ref{sec:verticalwaves}.

We consider a vertically stratified equilibrium with a
background shear flow $v_y=-3\Omega x/2$. Apart from $v_y$, all other
fluid quantities are constant in the horizontal direction.
We assume that the radiation and gas
exchange heat sufficiently rapidly that $T_{\rm rad}=T$ everywhere.
We also take the medium to be sufficiently optically
thick that the radiation pressure tensor ${\sf P}$ only
has diagonal elements $E/3$, and radiation transport is diffusive
with $\bar{\kappa}^{\rm R}\simeq\kappa_{\rm es}={\rm constant}$.
These thermodynamic assumptions are
excellent approximations in the regions of the simulations where the
acoustic waves are evident.  Finally, we approximate the equilibrium
magnetic field as being purely azimuthal and having only vertical gradients.

The equilibrium vertical structure is given by hydrostatic equilibrium,
\begin{equation}
{d\over dz}\left(P_{\rm therm}+{B^2\over8\pi}\right)=-\rho\Omega^2z,
\label{eq:hydrostatic}
\end{equation}
radiative equilibrium,
\begin{equation}
{dF\over dz}=Q
\end{equation}
and diffusive radiation transport,
\begin{equation}
F=-{c\over3\kappa_{\rm es}\rho}{dE\over dz}.
\end{equation}
Here $P_{\rm therm}\equiv p+E/3$ is the total thermal pressure.  Note that
we are including turbulence only insofar as it contributes
energy dissipation $Q$.

We now consider vertically propagating, longitudinal waves on this equilibrium.
The Eulerian perturbed fluid velocity $\delta v(z,t)=\partial\xi/\partial t$,
where $\xi$ is the vertical Lagrangian displacement of each fluid element.
The linearized continuity equation is
\begin{equation}
\Delta\rho=-\rho{\partial\xi\over\partial z},
\label{eq:pertcont}
\end{equation}
where here and elsewhere, $\Delta\equiv\delta+\xi(d/dz)$ refers to Lagrangian
perturbations.
The linearized vertical momentum equation is
\begin{equation}
\rho{\partial\delta v\over\partial t}=-{\partial \delta P_{\rm therm}
\over\partial z}-\delta\rho\Omega^2 z-{1\over4\pi}{\partial\over\partial z}
(B\delta B).
\label{eq:pertmomentum}
\end{equation}
The artificial viscosity term has been lost as it is nonlinear in the
perturbation velocity \citep{sn92}.
The linearized flux-freezing equation is
\begin{equation}
\delta B=-{\partial\over\partial z}(B\xi)
\label{eq:pertfluxfreezing}
\end{equation}
Finally, the linearized total internal energy equation may be written as
\begin{equation}
{\partial\over\partial t}\left({\Delta P_{\rm therm}\over P_{\rm therm}}\right)
-\Gamma_1{\partial\over\partial t}\left({\Delta\rho\over\rho}\right)=
{\Gamma_3-1\over P_{\rm therm}}\left(\delta\tilde{Q}-
{\partial\delta F\over\partial z}
\right),
\end{equation}
where $\Gamma_1$ and $\Gamma_3$ are the usual generalized adiabatic exponents
for a mixture of gas and radiation at the same temperature \citep{cha67},
both of which are close to $4/3$ for our radiation-dominated simulations.
This equation may be integrated in time to give
\begin{equation}
{\Delta P_{\rm therm}\over P_{\rm therm}}
-{\Gamma_1\Delta\rho\over\rho}=
{\Gamma_3-1\over P_{\rm therm}}\int^t\left(\delta\tilde{Q}-
{\partial\delta F\over\partial z}\right)dt^\prime.
\label{eq:pertinternalenergy}
\end{equation}

Assuming the mode periods are short enough that there is little energy exchange
between the modes and the turbulence, little turbulent dissipative heating,
and little radiative damping, we can for the moment set
$\delta\tilde{Q}=\delta F=0$
in the equations.  Assuming a time dependence $\propto\exp(-i\omega t)$,
equations (\ref{eq:pertcont})-(\ref{eq:pertfluxfreezing})
and (\ref{eq:pertinternalenergy}) may then be combined to give a single
equation in $\delta v$:
\begin{equation}
{\partial\over\partial z}\left[\left(\Gamma_1P_{\rm therm}+{B^2\over4\pi}\right)
{\partial\delta v\over\partial z}\right]+\rho(\omega^2-\Omega^2)\delta v=0.
\label{eq:waveequation}
\end{equation}

The vertical epicyclic mode is the simplest general solution to this equation,
with $\omega=\Omega$ and $\delta v$ being spatially constant.  The
breathing mode is the next simplest.  Provided $\Gamma_1$ is spatially
constant (which is true if gas or radiation dominates the thermal pressure),
and the magnetic energy density is much less than the thermal pressure
(an assumption that typically fails only in the low density surface layers),
then the hydrostatic equilibrium equation (\ref{eq:hydrostatic}) guarantees that
$\omega=(1+\Gamma_1)^{1/2}\Omega$ and $\delta v\propto z$ satisfy equation
(\ref{eq:waveequation}).

Like the breathing mode, higher order
modes are acoustic in nature, but their frequencies and eigenfunctions
depend more sensitively on the details of the equilibrium vertical structure
\citep{oka87,lub93,bla06}.

We now consider radiative damping of these modes.
Multiplying equation (\ref{eq:pertmomentum}) by $\delta v$, and then combining
with equations (\ref{eq:pertcont}), (\ref{eq:pertfluxfreezing}), and
(\ref{eq:pertinternalenergy}), we finally obtain after some algebra the
following wave energy conservation equation
\begin{eqnarray}
{\partial\over\partial t}\left[{1\over2}\rho\delta v^2+{1\over2}\Gamma_1
P_{\rm therm}
\left({\Delta\rho\over\rho}\right)^2+{1\over2}\rho\xi^2\Omega^2+{(\Delta B)^2
\over8\pi}\right]
+{\partial\over\partial z}\left[\delta v\Delta P_{\rm therm}+\delta v
\Delta\left({B^2\over8\pi}\right)\right]\cr
=-{(\Gamma_3-1)\over
\rho}{\partial\Delta\rho\over\partial t}\int^t\left(\delta\tilde{Q}-
{\partial\delta F\over\partial z}\right)dt^\prime.
\label{eq:waveenergy}
\end{eqnarray}

The right hand side of equation (\ref{eq:waveenergy}) represents sources
and sinks of wave energy due to perturbed turbulent dissipation and radiative
losses.  True microscopic viscosity and resistivity would also contribute
additional terms on the right hand side arising from additional terms in
the perturbed momentum and flux-freezing equations, but our numerical
dissipation scheme has no such terms.  In reality they would be present, but
likely even more important than these would be terms representing turbulent
fluctuations, including turbulent viscosity and resistivity, which we have
ignored here.  However, we believe the damping to
be dominated by radiative diffusion.

Assuming that the damping rate is small, we can estimate it by computing
a work integral for the modes (e.g. \citealt{cox80}).
Approximating all the perturbed quantities on the right hand side as being
perfectly periodic at the mode period $\Pi=2\pi/\omega_{\rm R}$, where
$\omega_{\rm R}\equiv{\rm Re}(\omega)$, we
can time-average equation (\ref{eq:waveenergy}) over this mode period and
integrate the right hand side by parts.  Then integrating over all height,
assuming negligible wave energy fluxes leaving the top and bottom boundaries,
we finally obtain for the exponential damping rate
$\Gamma\equiv{\rm Im}(\omega)$ of the mode amplitude
\begin{equation}
\Gamma=-{1\over2\Pi{\cal E}_{\rm mode}}\int dz(\Gamma_3-1)
\int_0^\Pi dt{\Delta\rho\over\rho}
\left(\delta\tilde{Q}-{\partial\delta F\over\partial z}\right),
\label{eq:workintegral}
\end{equation}
where
\begin{equation}
{\cal E}_{\rm mode}\equiv\int dz
\left[{1\over2}\rho\delta v^2+{1\over2}\Gamma_1 P_{\rm therm}
\left({\Delta\rho\over\rho}\right)^2+{1\over2}\rho\xi^2\Omega^2+{(\Delta B)^2
\over8\pi}\right]
\end{equation}
is the mode energy.

Because $\delta v$ and therefore $\xi$ are spatially constant for the
epicyclic mode, equation (\ref{eq:pertcont}) implies that $\Delta\rho=0$,
i.e. fluid elements preserve their density under this mode.  The work
integral therefore vanishes, and there is no damping of this mode.  This
is as one would expect:  the only way to change this mode is to exert
an external force on the system, or eject momentum through the vertical
boundaries.

The breathing mode and higher order acoustic modes are damped, however.
The linearized radiative diffusion equation is
\begin{equation}
\delta F=-F{\delta\rho\over\rho}-{c\over3\kappa_{\rm es}\rho}
{\partial\delta E\over\partial z}.
\label{eq:pertraddiff}
\end{equation}
If we consider a radiation dominated equilibrium, and neglect gas
and magnetic pressure contributions
to the mode dynamics, then equations (\ref{eq:pertmomentum}) and
(\ref{eq:pertraddiff}) give a simple relationship between the perturbed
radiative flux and the perturbed velocity,
\begin{equation}
\delta F={c\over\kappa_{\rm es}}{\partial\delta v\over\partial t}.
\end{equation}
Setting $\delta v=A_{\rm b}(z/H_P)\cos\omega_{\rm R}t$ for the breathing mode,
where $A_{\rm b}$ is the velocity amplitude at one pressure scale height
$H_P$ and $\omega_{\rm R}=(7/3)^{1/2}\Omega$, we find a radiative damping
rate of
\begin{equation}
\Gamma_{\rm rad}={c\Omega^2\over4\kappa_{\rm es}H_PE_0}\int dz,
\end{equation}
where $E_0$ is the midplane radiation energy density, and the integral
goes over the entire equilibrium, or at least that portion which is
optically thick.  Clearly this integral can be set equal to the pressure
scale height $H_P$ times some numerical constant.  For example, for an
$n=3$ isentropic polytrope, we obtain
\begin{equation}
\Gamma_{\rm rad}={315\over256}\left({c\Omega^2\over\kappa_{\rm es} E_0}
\right)\sim{c\Omega^2\over\kappa_{\rm es}E_0}.
\label{eq:gammaradappendix}
\end{equation}

The existence of this damping produces nonzero time-averages in the wave
energy fluxes because it changes the phase difference between the velocity
perturbation $\delta v$ and the pressure perturbations $\Delta P_{\rm therm}$
and $\Delta(B^2/8\pi)$.  In the absence of an exact solution to the
non-adiabatic
eigenfunctions, we compute these pressure perturbations using the adiabatic
equations.  Setting $\delta v=A_{\rm b}(z/H_P)e^{-\Gamma t}\cos\omega_{\rm R}t$,
then the instantaneous energy fluxes are
\begin{equation}
\delta v\Delta P_{\rm therm}\simeq\delta v{\Delta E\over3}=
-{4EzA_{\rm b}^2\over9\omega_{\rm R}H_P^2}
e^{-2\Gamma t}\left(\cos\omega_{\rm R}t\sin\omega_{\rm R}t-
{\Gamma\over\omega_{\rm R}}\cos^2\omega_{\rm R}t\right)+
{\cal O}\left({\Gamma^2\over\omega_{\rm R}^2}\right)
\end{equation}
and
\begin{equation}
\delta v\Delta\left({B^2\over8\pi}\right)=-{B^2zA_{\rm b}^2\over4\pi
\omega_{\rm R}H_P^2}
e^{-2\Gamma t}\left(\cos\omega_{\rm R}t\sin\omega_{\rm R}t-
{\Gamma\over\omega_{\rm R}}\cos^2\omega_{\rm R}t\right)+
{\cal O}\left({\Gamma^2\over\omega_{\rm R}^2}\right).
\end{equation}
The first term in each of these expressions periodically changes sign, and
a time-average of this term over one mode period results in positive or
negative values depending on the phase of the time-interval chosen.  This term
therefore produces no secular energy flux.  The second term does, and the
resulting time-averaged fluxes are, to lowest order in
$\Gamma/\omega_{\rm R}$,
\begin{equation}
<\delta v\Delta P_{\rm therm}>={2Ez\Gamma A_{\rm b}^2\over21\Omega^2H_P^2}
\label{eq:dvdpthappendix}
\end{equation}
and
\begin{equation}
\left<\delta v\Delta\left({B^2\over8\pi}\right)\right>={3B^2z\Gamma
A_{\rm b}^2\over56\pi\Omega^2H_P^2}.
\label{eq:dvdpmagappendix}
\end{equation}
Note that these two equations are completely independent of whether the
decay rate $\Gamma$ was due to radiation damping or some other process.


\begin{thebibliography}{}
\bibitem[Acheson(1979)]{ach79}Acheson, D. J. 1979, Solar Phys., 62, 23
\bibitem[Agol \& Krolik(1998)]{ago98}Agol, E., \& Krolik, J. 1998, ApJ, 507,
304
\bibitem[Balbus \& Hawley(1991)]{bal91}Balbus, S. A., \& Hawley, J. F. 1991,
ApJ, 376, 214
\bibitem[Balbus \& Hawley(1998)]{bal98}Balbus, S. A., \& Hawley, J. F. 1998,
Rev. Mod. Phys., 70, 1
\bibitem[Balbus \& Henri(2008)]{bal08}Balbus, S. A., \& Henri, P. 2008,
ApJ, 674, 408
\bibitem[Begelman(2001)]{beg01}Begelman, M. C. 2001, ApJ, 551, 897
\bibitem[Begelman(2006)]{beg06}Begelman, M. C. 2006, ApJ, 643, 1065
\bibitem[Bisnovatyi-Kogan \& Blinnikov(1977)]{bis77}Bisnovatyi-Kogan, G. S.,
\& Blinnikov, S. I. 1977, A\&A, 59, 111
\bibitem[Blaes, Arras, \& Fragile(2006)]{bla06}Blaes, O. M., Arras, P.,
\& Fragile, P. C. 2006, MNRAS, 369, 1235
\bibitem[Blaes, Hirose, \& Krolik(2007)]{bla07}Blaes, O., Hirose, S., \& Krolik,
J. H. 2007, ApJ, 664, 1057
\bibitem[Blaes \& Socrates(2003)]{bla03}Blaes, O., \& Socrates, A. 2003, ApJ, 5
96, 509
\bibitem[Brandenburg et al.(1995)]{bra95}Brandenburg, A., Nordlund, {\AA}.,
Stein, R.F. \& Torkelsson, U. 1995, ApJ, 446, 741
\bibitem[Chandrasekhar(1967)]{cha67}Chandrasekhar, S. 1967, An Introduction
to the Study of Stellar Structure (New York:  Dover)
\bibitem[Cox(1980)]{cox80}Cox, J. P. 1980, Theory of Stellar Pulsation
(Princeton: Princeton University Press)
\bibitem[Davis et al.(2009)]{dav10}Davis, S. W., Stone, J. M., \& Pessah,
M. E. 2010, ApJ, 713, 52
\bibitem[Hawley, Gammie, \& Balbus(1995)]{haw95} Hawley, J. F., Gammie, C. F.,
\& Balbus, S. A. 1995, ApJ, 440, 742
\bibitem[Fromang et al.(2007)]{fplh07}Fromang, S., Papaloizou, J., Lesur, G.,
\& Heinemann, T. 2007, A\&A, 476, 1123
\bibitem[Galeev, Rosner \& Vaiana(1979)]{gal79}Galeev, A. A., Rosner, R., \&
Vaiana, G. S. 1979, ApJ, 229, 318
\bibitem[Gammie(1998)]{gam98}Gammie, C. F. 1998, MNRAS, 297, 929
\bibitem[Gilman(1970)]{gil70}Gilman, P. A. 1970, ApJ, 162, 1019
\bibitem[Gressel(2010)]{gre10}Gressel, O. 2010, MNRAS, 405, 41
\bibitem[Hawley \& Balbus(1991)]{haw91}Hawley, J. F., \& Balbus, S. A. 1991,
ApJ, 376, 223
\bibitem[Heinemann \& Papaloizou(2009a)]{hei09a}Heinemann, T., \& Papaloizou,
J. C. B. 2009a, MNRAS, 397, 52
\bibitem[Heinemann \& Papaloizou(2009b)]{hei09b}Heinemann, T., \& Papaloizou,
J. C. B. 2009b, MNRAS, 397, 64
\bibitem[Hirose, Krolik, \& Stone(2006)]{hir06}Hirose, S., Krolik, J. H.,
\& Stone, J. M. 2006, ApJ, 640, 901
\bibitem[Hirose, Blaes, \& Krolik(2009)]{hirbk09}Hirose, S., Blaes, O., \&
Krolik, J. H. 2009, ApJ, 704, 781
\bibitem[Hirose, Krolik, \& Blaes(2009)]{hir09}Hirose, S., Krolik, J. H., \&
Blaes, O. 2009, ApJ, 691, 16
\bibitem[Johansen, Youdin, \& Klahr(2009)]{joh09}Johansen, A., Youdin, A.,
\& Klahr, H. 2009, ApJ, 697, 1269
\bibitem[Krolik, Hirose, \& Blaes(2007)]{kro07}Krolik, J. H., Hirose, S.,
\& Blaes, O.  2007, ApJ, 664, 1045
\bibitem[Lesur \& Longaretti(2007)]{les07}Lesur, G., \& Longaretti, P.-Y.
2007, MNRAS, 378, 1471
\bibitem[Lightman \& Eardley(1974)]{le74}Lightman, A. P., \& Eardley, D. M.
1974, ApJ, 187, L1
\bibitem[Lubow \& Pringle(1993)]{lub93}Lubow, S. H., \& Pringle, J. E. 1993,
ApJ, 409, 360
\bibitem[Miller \& Stone(2000)]{mil00}Miller, K. A., \&
Stone, J. M.  2000, ApJ, 534, 398
\bibitem[Newcomb(1961)]{new61}Newcomb, W. A. 1961, Phys. Fluids, 4, 391
\bibitem[Novikov \& Thorne(1973)]{nov73}Novikov, I. D., \& Thorne, K. S. 1973,
in Black Holes, ed. C. DeWitt-Morette \& B. S. DeWitt (New York:  Gordon
\& Breach), 343
\bibitem[Ohsuga et al.(2009)]{ohs09}Ohsuga, K., Mineshige, S., Mori, M.,
\& Kato, Y. 2009, PASJ, 61, L7
\bibitem[Okazaki, Kato \& Fukue(1987)]{oka87}Okazaki, A. T., Kato, S.,
\& Fukue, J. 1987, PASJ, 39, 457
\bibitem[O'Neill et~al.(2010)]{one10}O'Neill, S.M., Reynolds, C.S., Coleman,
M.C. \& Sorathia, K. 2010, arXiv:1009.1882
\bibitem[Sakimoto \& Coroniti(1981)]{sak81}Sakimoto, P. J., \& Coroniti, F. V.
1981, ApJ, 247, 19
\bibitem[Sakimoto \& Coroniti(1989)]{sak89}Sakimoto, P. J., \& Coroniti, F. V.
1989, ApJ, 342, 49
\bibitem[Shakura \& Sunyaev(1973)]{sha73}Shakura, N. I., \& Sunyaev, R. A.
1973, A\&A, 24, 337
\bibitem[Shakura \& Sunyaev(1976)]{sha76}Shakura, N. I., \& Sunyaev, R. A.
1976, MNRAS, 175, 613
\bibitem[Shakura, Sunyaev \& Zilitinkevich(1978)]{sha78}Shakura, N. I.,
Sunyaev, R. A., \& Zilitinkevich, S. S. 1978, A\&A, 62, 179
\bibitem[Shi, Krolik, \& Hirose(2010)]{shi10}Shi, J., Krolik, J. H., \&
Hirose, S.  2010, ApJ, 708, 1716
\bibitem[Shibazaki \& H\=oshi(1975)]{shi75}Shibazaki, N., \& H\=oshi, R. 1975,
Prog. Theor. Phys., 54, 706
\bibitem[Silk(1967)]{sil67}Silk, J. 1967, Nature, 215, 1155
\bibitem[Silk(1968)]{sil68}Silk, J. 1968, ApJ, 151, 459
\bibitem[Simon \& Hawley(2009)]{sim09}Simon, J. B., \& Hawley, J. F. 2009, ApJ,
707, 833
\bibitem[Stella \& Rosner(1984)]{ste84}Stella, L., \& Rosner, R. 1984,
ApJ, 277, 312
\bibitem[Stone et~al.(1996)]{sto96}Stone, J. M., Hawley, J. F., Gammie,
C. F., \& Balbus, S. A. 1996, ApJ, 463, 656
\bibitem[Stone \& Norman(1992)]{sn92}Stone, J. M., \& Norman, M. L. 1992, ApJS, 80, 753
\bibitem[Stone, Mihalas \& Norman(1992)]{sto92}Stone, J. M., Mihalas, D.,
\& Norman, M. L. 1992, ApJS, 80, 819
\bibitem[Syunyaev \& Shakura(1975)]{sun75}Syunyaev, R. A.,
\& Shakura, N. I. 1975, SvAL, 1, 158
\bibitem[Turner(2004)]{tur04}Turner, N. J., 2004, ApJ, 605, L45
\bibitem[Turner, Stone \& Sano(2002)]{tur02}Turner, N. J., Stone, J. M., \&
Sano, T. 2002, ApJ, 566, 148
\bibitem[Turner et~al.(2003)]{tur03}Turner, N. J., Stone, J.M., Krolik, J.H.,
\& Sano, T. 2003, ApJ, 593, 992
\bibitem[Turner et al.(2005)]{tur05}Turner, N. J., Blaes, O. M., Socrates,
A., Begelman, M. C., \& Davis, S. W. 2005, ApJ, 624, 267
\end{thebibliography}
\end{document}